\documentclass[aps,pre,longbibliography,showpacs,amsmath,amssymb]{revtex4-1}
\usepackage{amsmath,amssymb,latexsym,epsfig,graphics,epsf}
\usepackage{hyperref}
\usepackage{graphics}
\usepackage{float}
\usepackage{footnote}
\usepackage[utf8]{inputenc}
\usepackage{adjustbox}
\usepackage{blindtext}
\usepackage[T1]{fontenc}
\usepackage{lpic}
\usepackage{tabularx,booktabs}
\usepackage[dvipsnames]{xcolor}

\newcommand{\be}{\begin{equation}}
\newcommand{\ee}{\end{equation}}
\newcommand{\fig}[1]{Fig.~\ref{#1}}
\newcommand{\Fig}[1]{Figure~\ref{#1}}
\newcommand{\eq}[1]{Eq.~(\ref{#1})}

\newcommand{\Sex}{{S}_{\rm ex}}

\newcommand{\bR}{\bold R}

\newcommand{\br}{\bold r}

\newcommand{\tbr}{{\tilde{\bold r}}}

\begin{document}
	\title{Scaling Properties of Liquid Dynamics Predicted from a Single Configuration: Pseudoisomorphs for Harmonic-Bonded Molecules}
	\date{\today}
	\author{Zahraa Sheydaafar${\rm ^a}$}
	\author{Jeppe C. Dyre${\rm ^a}$}
	\author{Thomas B. Schr{\o}der${\rm ^a}$}\email{tbs@ruc.dk}
	\affiliation{${\rm ^a}$Glass and Time, IMFUFA, Department of Science and Environment, Roskilde University, P.O. Box 260, DK-4000 Roskilde, Denmark}

\begin{abstract}
Isomorphs are curves in the thermodynamic phase diagram of invariant excess entropy, structure, and dynamics, while pseudoisomorphs are curves of invariant structure and dynamics, but not of the excess entropy. The latter curves have been shown to exist in molecular models with flexible bonds [A. E. Olsen et al., \textit{J. Chem. Phys.} \textbf{2016}, \textit{145}, 241103 (2016)]. We here present three force-based methods to trace out pseudoisomorphs based on a single configuration and test them on the asymmetric dumbbell and 10-bead Lennard-Jones chain models with bonds modeled as harmonic springs. The three methods are based on requiring that particle forces, center-of-mass forces, and torques, respectively, are invariant in reduced units. For each of the two investigated models we identify a method that works well for tracing out pseudoisomorphs, but these methods are not the same. Overall, it appears that the more internal degrees of freedom there are in the molecule studied, the less they appear to affect the gross dynamical behavior. Moreover, the ``internal'' degrees of freedom (including rotation) do not appear to significantly affect the scaling behavior of the dynamical/transport coefficients provided some 'quenching' is performed. 
\end{abstract}

\maketitle

\section{Introduction}
Isomorphs are curves of constant excess entropy in the thermodynamic phase diagram along which structure and dynamics are invariant to a good approximation \cite{IV}. We remind that the excess entropy $\Sex$ is the entropy minus that of an ideal gas at the same density and temperature \cite{han13}. Systems with isomorphs, termed R-simple \cite{mal13,fle14,pra14,sch14,hey15,khr16,kas23}, are characterized by a strong correlation between the canonical-ensemble equilibrium fluctuations of potential energy $U$ and virial $W$ as quantified by the Pearson correlation coefficient (sharp brackets denote  canonical averages and $\Delta$ the deviation from the mean),

\be\label{R}
R = \dfrac{\langle \Delta W \Delta U \rangle}{\sqrt{\langle (\Delta W)^2 \rangle \langle (\Delta U)^2 \rangle}}\,.
\ee
The criterion for being R-simple is $R>0.9$ \cite{IV}, but even systems with somewhat lower R values may have good isomorphs \cite{ing12b}.  

Isomorph invariance of structure and dynamics refers to the use of units where the energy unit is $e_0\equiv k_BT$, the length unit is $l_0\equiv\rho^{-1/3}$ in which $\rho$ is the particle number density, i.e., the density of atoms, and the time unit is $t_0\equiv \rho^{-1/3}\sqrt{m/k_BT}$ in which $m$ is a particle mass. Quantities made dimensionless by reference to this unit system are termed ``reduced'' and marked by a tilde; for instance the reduced particle position $\br$ is given by $\tbr\equiv\br/l_0=\rho^{1/3}\br$. 

The existence of isomorphs for a given system means that the thermodynamic phase diagram is essentially one-dimensional in regard to structure and dynamics. Thus if one imagines filming how the molecules move, two state points on the same isomorph would give rise to (almost) the same movie -- except for an overall scaling of space and time. This provides an significant simplification of the physics, basically saying that the change of dynamics induced by a pressure increase may be counteracted by increasing at the same time the temperature. Many systems, however, do not have isomorphs in the original sense of the word as lines of constant excess entropy \cite{IV}, which raises the question: Can some systems still have lines of (approximately) invariant structure and dynamics? This question motives the below presented investigation of two molecular models.

Because $\Sex$ is the part of entropy that refers to particle positions, isomorphs are configurational adiabats. These can be traced out in the thermodynamic phase diagram by using the generally valid statistical-mechanical relation \cite{IV}:

\be\label{gamma}
\gamma \equiv  \left( \frac{\partial \ln T}{\partial \ln \rho}\right)_{\mathrm{S_{ex}}} = 
\frac{\langle \Delta U \Delta W \rangle}{\langle (\Delta U)^2 \rangle}\,. 
\ee 
Evaluating the right-hand side by equilibrium $NVT$ simulations, an isomorph is traced out by solving the differential equation Eq.~(\ref{gamma}) numerically using, e.g., the Euler or Runge-Kutta methods \cite{att21}.

\begin{figure}[htbp!]
   \includegraphics[width=8cm]{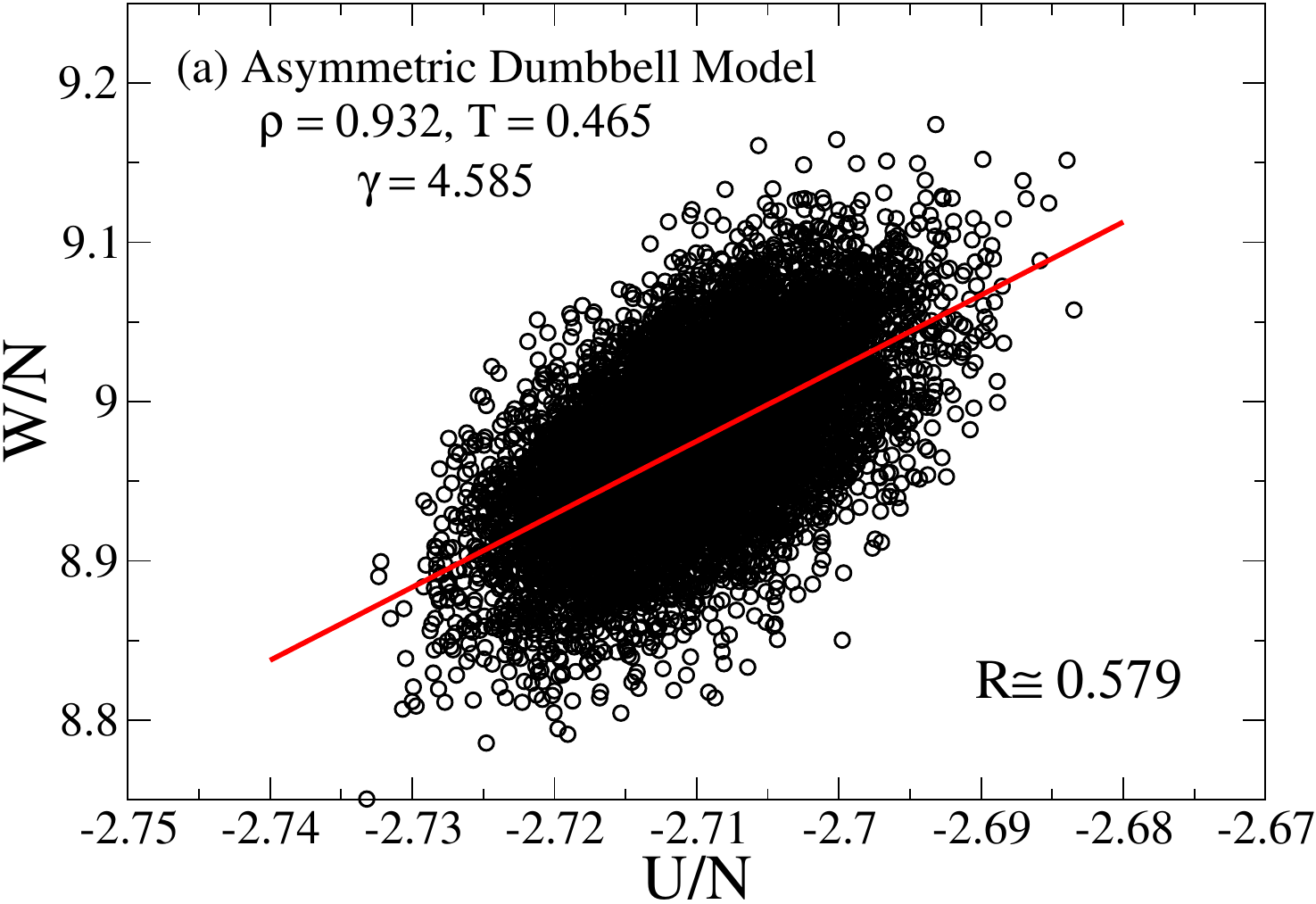}
   \includegraphics[width=8cm]{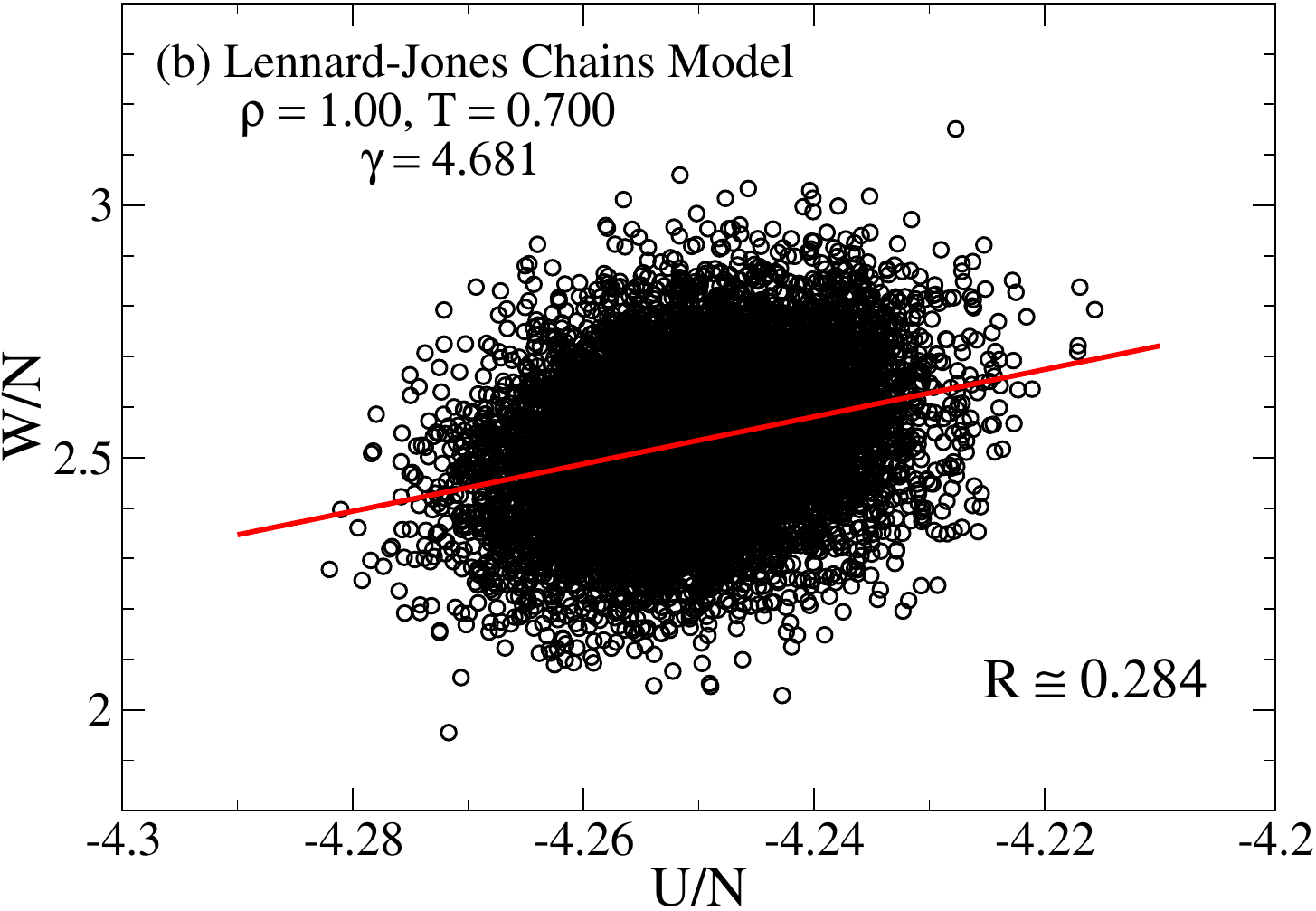}
	\caption{Scatter plots of equilibrium fluctuations of the virial and the potential energy.
		(a) Results for the asymmetric dumbbell model with harmonic bonds and
		(b) for the 10-bead flexible Lennard-Jones chain model with harmonic bonds.
		For both models the correlation coefficient $R$ is much smaller than the criterion for an R-simple system, $R>0.9$ \cite{IV}.
        \label{FIG:ASD_LJC_Spr}
	}
\end{figure}

Isomorphs have been identified and validated for both atomic \cite{IV,boh12,EXPII,hey19,att21,cas21} and molecular systems \cite{ing12b,vel14}, and isomorph-theory predictions have also been verified in experiments on glass-forming van der Waals molecular liquids \cite{xia15,han18}. In molecular systems, isomorphs are found when bonds are modeled as constraints \cite{ing12b,vel14}, but not when the bonds are flexible \cite{vel15a,ols16,kas23}. As an example, \fig{FIG:ASD_LJC_Spr} shows scatter plots of virial versus potential energy for the asymmetric dumbbell (ASD) and 10-bead Lennard-Jones chain (LJC) models, both with harmonic springs as commonly used in simulations \cite{kop22,kas23}. Neither model is R-simple; the virial potential-energy correlation coefficients are $0.579$ and $0.284$, respectively. As expected, these models do not have isomorphs (data not shown), i.e., structure and dynamics are not invariant along the curves of constant $\Sex$. Nevertheless, it has been found that both models have curves in the phase diagram of invariant structure and dynamics; termed ``pseudoisomorphs'' \cite{vel14,kas23}.

\begin{figure}[htbp!]
 	\centering
 	\includegraphics[width=14cm]{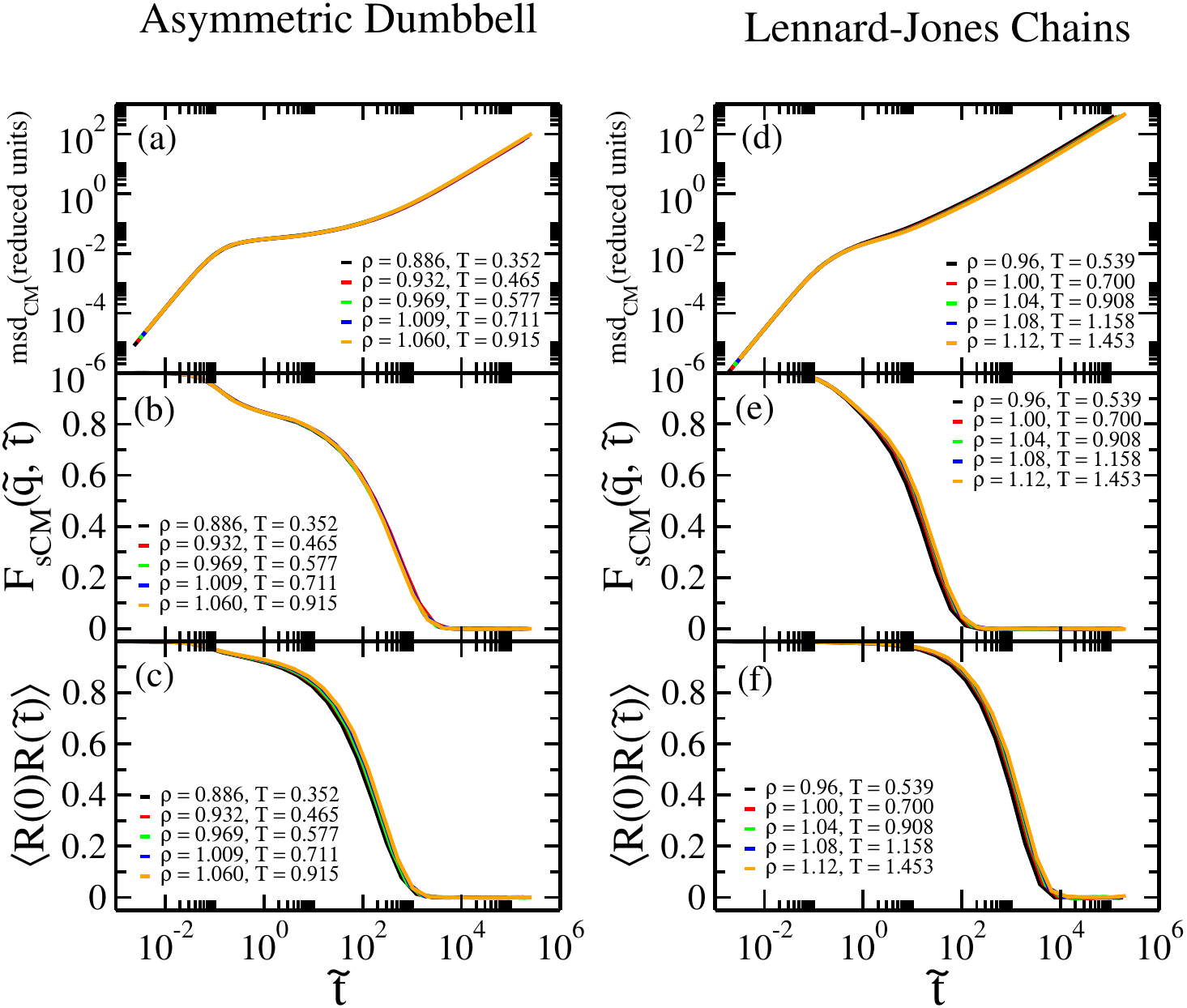}	
 	\caption{\label{FIG:Pseudo-LJC-ASD}Invariance of the dynamics along state points determined by Olsen      \textit{et al.} \cite{ols16} to define pseudoisomorphs. 
            (a), (b), (c) show results for the asymmetric dumbbell model with harmonic bonds; 
            (d), (e), (f) show results for the 10-bead Lennard-Jones chain model with harmonic bonds. 
 		(a) and (d) report the reduced mean-square displacement of the center of mass as a function of reduced time.
 		(b) and (e) report the center-of-mass incoherent intermediate scattering function as a function of reduced time at the wave-vector corresponding to the first peak of the static structure factor.
 		(c) and (f) report the normalized end-to-end vector time-autocorrelation function, probing the decay of molecular orientation.
            }
\end{figure}

Since pseudoisomorphs are not configurational adiabats, Eq.~(\ref{gamma}) cannot be used to identify such curves in the phase diagram. In 2016 Olsen \textit{et al.} presented a method for tracing out pseudoisomorphs involving the following steps \cite{ols16}: 
1) An equilibrium configuration is quenched to the nearest local minimum in the high-dimensional potential energy landscape, the so-called inherent state \cite{sti83}; 
2) the Hessian matrix is diagonalized to find the vibrational spectrum of the inherent state; 
3) the high-frequency part of the spectrum (related to the springs) is identified; 
4) the scaling properties upon a density change of the remaining part of the spectrum is used to identify the pseudoisomorph. 
This method is very computationally demanding, but works well. Olsen \textit{et al.} demonstrated that the it traces out pseudoisomorphs for the ASD and the flexible LJC models demonstrating good, though not perfect invariance of the dynamics probed via the incoherent intermediate scattering function. For comparison with later results, \fig{FIG:Pseudo-LJC-ASD} shows the dynamics probed via the mean-square displacement, incoherent intermediate scattering function, and orientational time-autocorrelation function, for the state points identified in \cite{ols16}.

The present paper proposes simpler methods for generating pseudoisomorphs based on the scaling properties of the forces of a single configuration \cite{sch22}. This idea has been shown to work well not only for atomic systems like the Kob-Andersen binary Lennard-Jones mixture, but also for molecular models like the ASD model and the Lewis-Wahnstrom OTP model with rigid bonds \cite{ing12b,kop20,sch22,she23}.

\section{Models and Simulation details}

The ASD model is a toy model of toluene \cite{vra01,gal07,III,cho10a,cho10b,ing12b,fra17,san18,dom20}. We simulated a system of 5000 ASD molecules defined as two different-sized spheres, a large (A) and a small (B) one \cite{vra01,sch09,mil13}. The spheres interact via Lennard-Jones (LJ) potentials, and in the units defined by the large sphere ($\sigma_{AA} \equiv 1$, $\epsilon_{AA} \equiv 1$, and $m_A \equiv 1$) the other LJ-parameters are  $\sigma_{AB} = 0.894$, $\sigma_{BB} = 0.788$, $\epsilon_{AB} = 0.342$, $\epsilon_{BB} = 0.117$, $m_B = 0.195$. Bonds are modeled as harmonic springs with equilibrium length $0.584$ and spring constant $k=3000$.

The LJC model is a generic coarse-grained polymer model \cite{ben98,aic03,puo11,sha13}. We simulated 1000 10-bead LJC molecules. Non-bonded particles interact via a standard LJ potential, cutting and shifting the forces at $2.5\sigma$. All particles are of same type and the potential parameters are set to unity: $\sigma = 1, \epsilon = 1$. Bonds are modeled as harmonic springs with equilibrium length $1$ and spring constant $k=3000$.

All Molecular Dynamics simulations were performed in the $NVT$ ensemble with a Nose-Hoover thermostat in periodic boundary conditions, using RUMD that is an open-source molecular dynamics package optimized for GPU computing  \cite{RUMD} (\url{http://rumd.org}).

\section{Identifying pseudoisomorphs via force-based methods}

Single-configuration force-based methods for generating isomorphs were introduced recently \cite{sch22,she23}. The idea is the following. Given a configuration $\mathbf{R}_1$ at state point $(\rho_1, T_1)$, a uniform scaling to density $\rho_2$ is performed leading to $\mathbf{R}_{2} = \left(\rho_1/\rho_2 \right)^{1/3} \mathbf{R}_{1}$. For molecules, two variants of this scaling can be applied: ``center-of-mass scaling'' where the molecular center-of-masses  are scaled while all orientations and internal degrees of freedom are kept fixed, or ``atomic scaling'' where a uniform scaling is applied to all atoms thus modifying also the intramolecular bond lengths. After scaling a configuration, the forces associated with the two configurations, $\mathbf{F}(\mathbf{R}_1)$ and $\mathbf{F}(\mathbf{R}_2)$, are compared ($\mathbf F$ is the long vector of all forces). The temperature $T_2$ at the density $\rho_2$ is identified from the condition of invariant reduced forces. Specifically, $|\mathbf{\tilde{F}}(\mathbf{R}_1)| = |\mathbf{\tilde{F}}(\mathbf{R}_2)|$ implies \cite{sch22}

\be\label{eq:pre_temp}
T_2 = \frac{|\mathbf{F}(\mathbf{R}_2)|}{|\mathbf{F}(\mathbf{R}_1)|} \left(\frac{\rho_1}{\rho_2}\right)
^{1/3} T_1\,.
\ee
If the two force vectors are parallel, this leads to the forces being identical in reduced units, $\mathbf{\tilde{F}}(\mathbf{R}_1) = \mathbf{\tilde{F}}(\mathbf{R}_2)\,$. Assuming this is representative of all relevant configurations, it follows that structure and dynamics are invariant in reduced units because the same reduced-unit equation of motion applies at the two state points \cite{sch14,dyr18a}. We do not expect the force vectors before and scaling to be perfectly parallel, however, but \eq{eq:pre_temp} can still be used. Reference \cite{sch22} proposed that for force-based method to work well, both the Pearson and Spearman correlation coefficients of the force components should be larger than 0.95 (the latter is defined as the Pearson correlation coefficient of the rank of the data). 

Different variants of the method are arrived at by different interpretations of what exactly $\mathbf{F}(\mathbf{R})$ represents, e.g., the forces on all the atoms or just on the center-of-mass forces. We consider below also a variant based on invariance of torques in reduced units, $\boldsymbol{\tilde{\tau}_1} = \boldsymbol{\tilde{\tau}_2}$, which leads to

\begin{equation}\label{eq:pre_tor}
	T_2 = \frac{\boldsymbol{|\tau_2|}}{\boldsymbol{|\tau_1|}} T_1\,.
\end{equation}

\begin{figure}[htbp!]
	\centering
	\includegraphics[width=7cm]{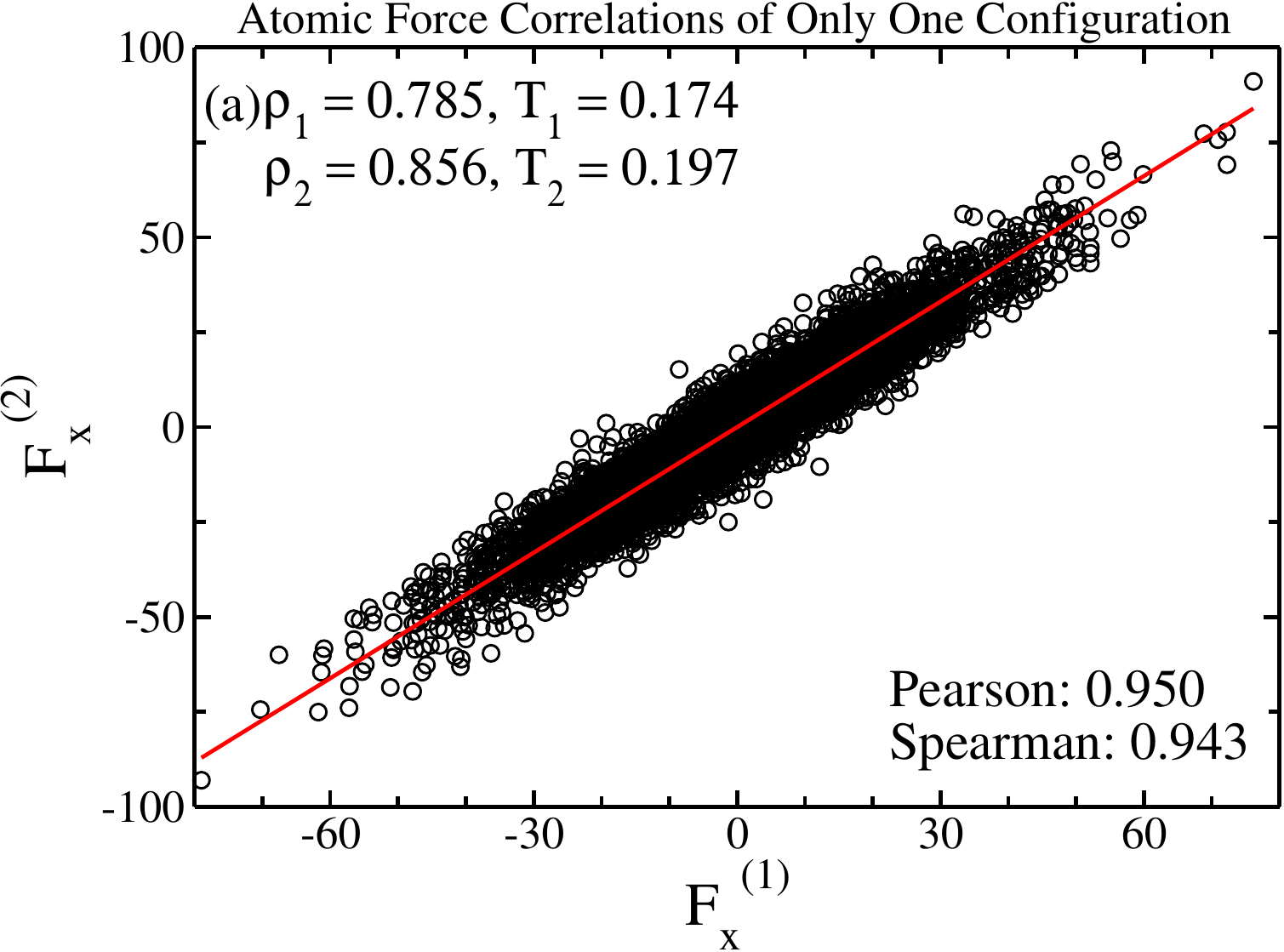}
	\includegraphics[width=7cm]{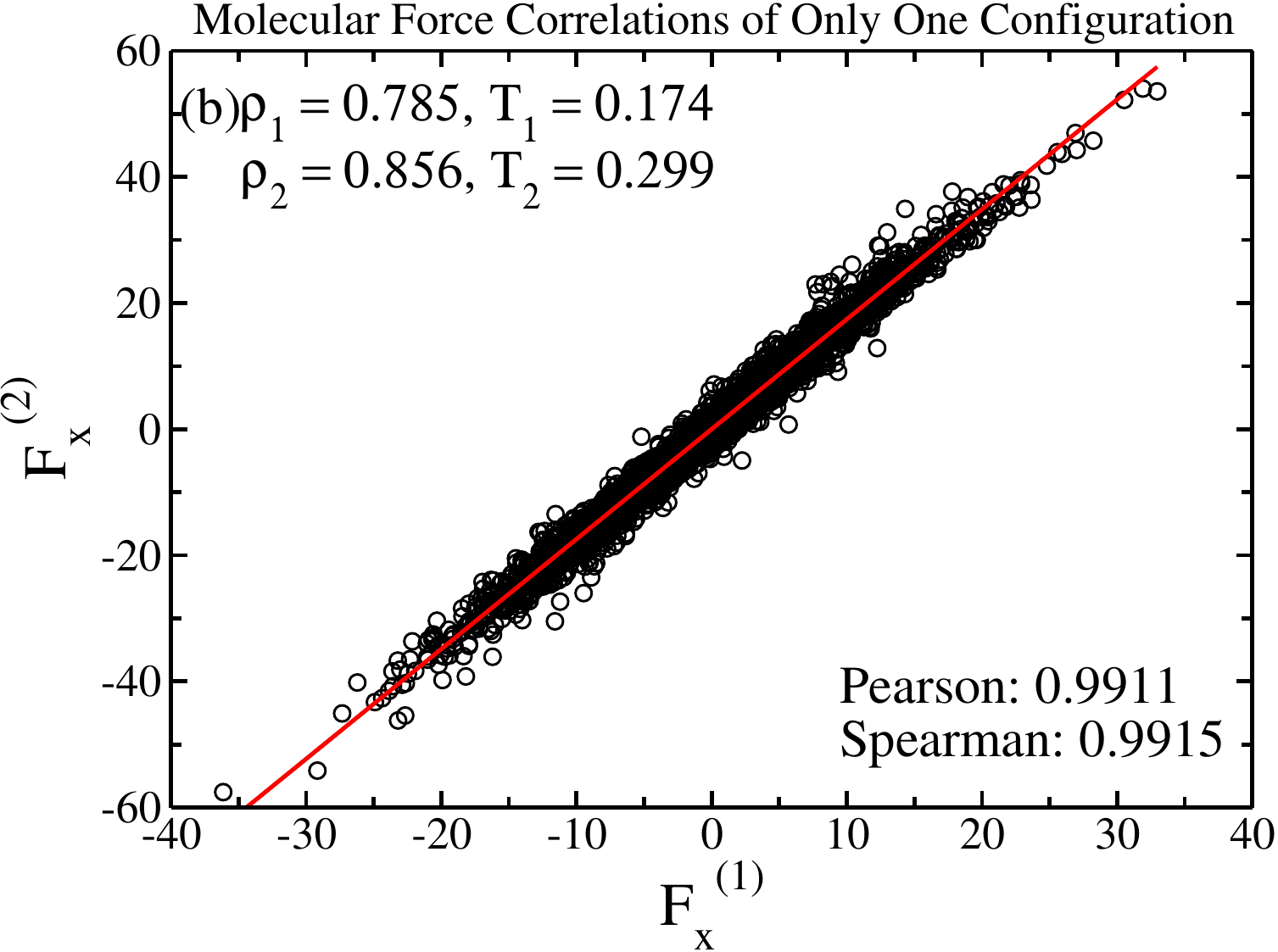}	
	\includegraphics[width=7cm]{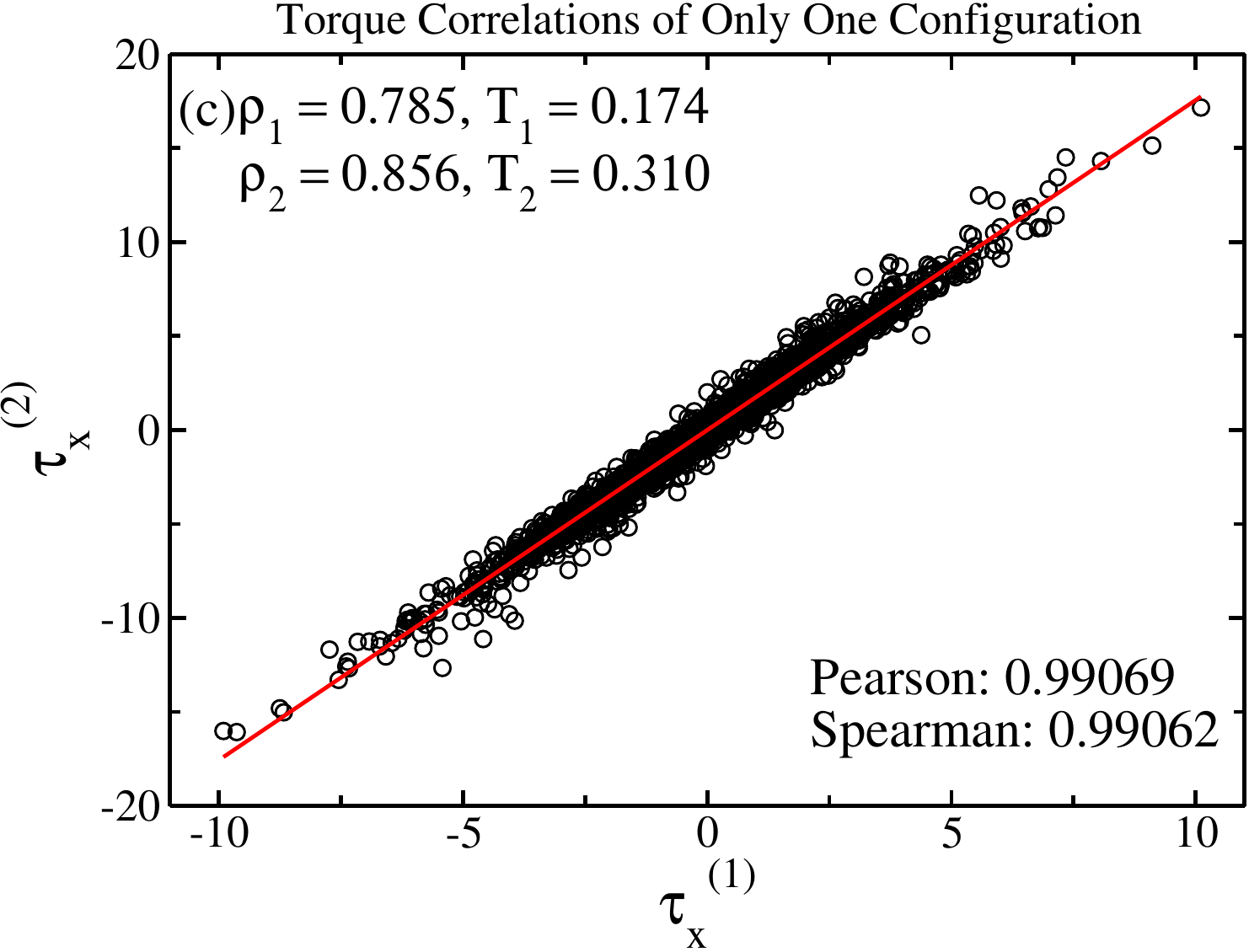}
	\caption{\label{cor_F_ASD_low} Force and torque correlations of a single configuration of the harmonic ASD model using center-of-mass scaling. Each point represents scaled and unscaled forces/torques.
	(a) Atomic-force method. The figure shows the x-coordinates of the reduced particle forces plotted against the same quantities of the uniformly scaled configuration. The temperature $T_2=0.197$ is identified by means of \eq{eq:pre_temp}. 
	(b) Molecular-force method. The figure shows the center-of-mass ``molecular'' forces before and after scaling, which have no contributions from the springs. In this case $T_2=0.299$.
	(c)  Torque method. The figure shows the correlation between the torques of the molecules before and after scaling ($T_2=0.310$).
}
\end{figure}

\section{Results for the ASD model}

The three methods are applied to the ASD model in \fig{cor_F_ASD_low}, using center-of-mass scaling, i.e., $\bR_{2,cm}=(\rho_1/\rho_2)^{1/3}\bR_{1,cm}$. The state point $(\rho_1, T_1) = (0.785, 0.174)$ is used as reference and $\rho_2=0.856$, i.e., a 9\% density increase is considered. \Fig{cor_F_ASD_low}(a) shows a scatter plot of the atomic-force components before and after scaling for a single configuration; here $T_2 = 0.197$ is found by applying atomic forces in \eq{eq:pre_temp}. \Fig{cor_F_ASD_low}(b) shows a similar plot based on the ``molecular'' forces, i.e., the center-of-mass forces defined as the sum of all forces on the atoms of a given molecule. In this case the quite different $T_2=0.299$ is arrived at and a significantly better correlation is obtained. Finally, \fig{cor_F_ASD_low}(c) shows the torque correlations of molecules of unscaled and scaled configurations. The correlation is here not far from that of the molecular forces; using \eq{eq:pre_tor} gives $T_2 = 0.310$.

 \begin{figure}[htbp!]
	\centering
	\includegraphics[width=14cm]{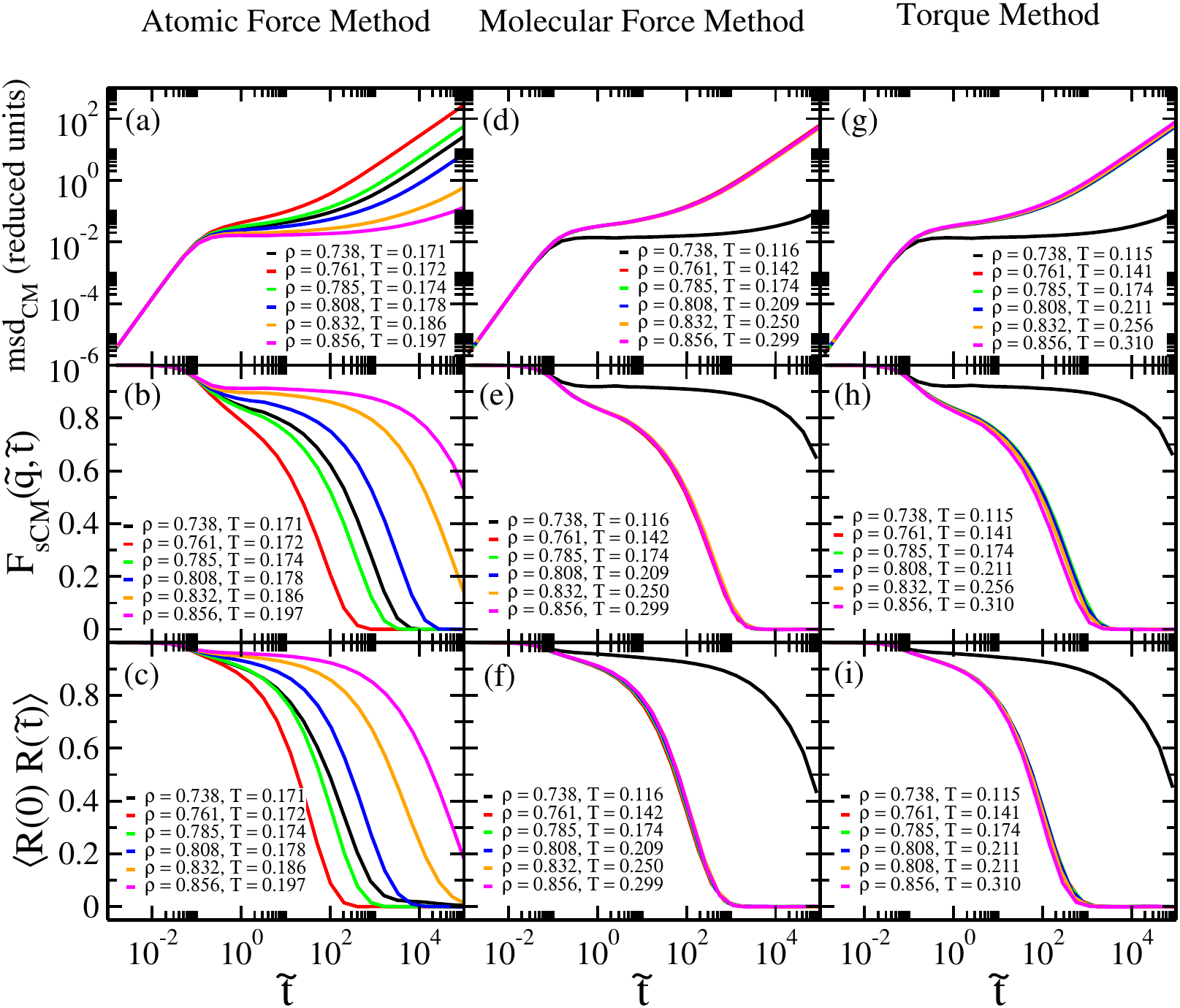}
	\caption{\label{ch3:FIG:Dyn_ASD_Low_Den} Testing for invariance of the reduced dynamics of the harmonic-spring ASD model using the three methods generating pseudoisomorphs by scaling a single configuration of the reference state point $(\rho_1, T_1) = (0.785, 0.174)$. Each method investigates the reduced center-of-mass mean-square displacement (upper figures), center-of-mass incoherent intermediate scattering function at the wave vectors corresponding to the first peak of the static structure factor (middle figures), and directional time-autocorrelation function probed via the autocorrelation function of the normalized bond vectors (bottom figures).
	(a), (b), (c) show results of the atomic-force method requiring invariant reduced forces between all atoms, i.e., including the harmonic bond contributions into \eq{eq:pre_temp}.
	(d), (e), (f) show results of the molecular-force method requiring invariant reduced center-of-mass forces between the molecules.
	(g), (h), (i) show results of the torque method requiring invariant reduced torques on the molecules (\eq{eq:pre_tor}).
	}
\end{figure}

For each of the three  methods, \fig{ch3:FIG:Dyn_ASD_Low_Den} compares results for the (same) dynamics at the state points, which the different methods propose to give identical dynamics. While the methods in principle require only a single configuration, for carefully comparing the methods, $T_2$-values were obtained by averaging over 195 independent configurations. We find that the molecular-force method gives almost invariant dynamics, except at the lowest density (black curve) where negative pressure and phase separation is observed. The torque method gives similar, though slightly inferior results, while the atomic force method does not work well.

\section{Results for the LJC model}

\begin{figure}[htbp!]
\centering
    \includegraphics[width=1\textwidth]{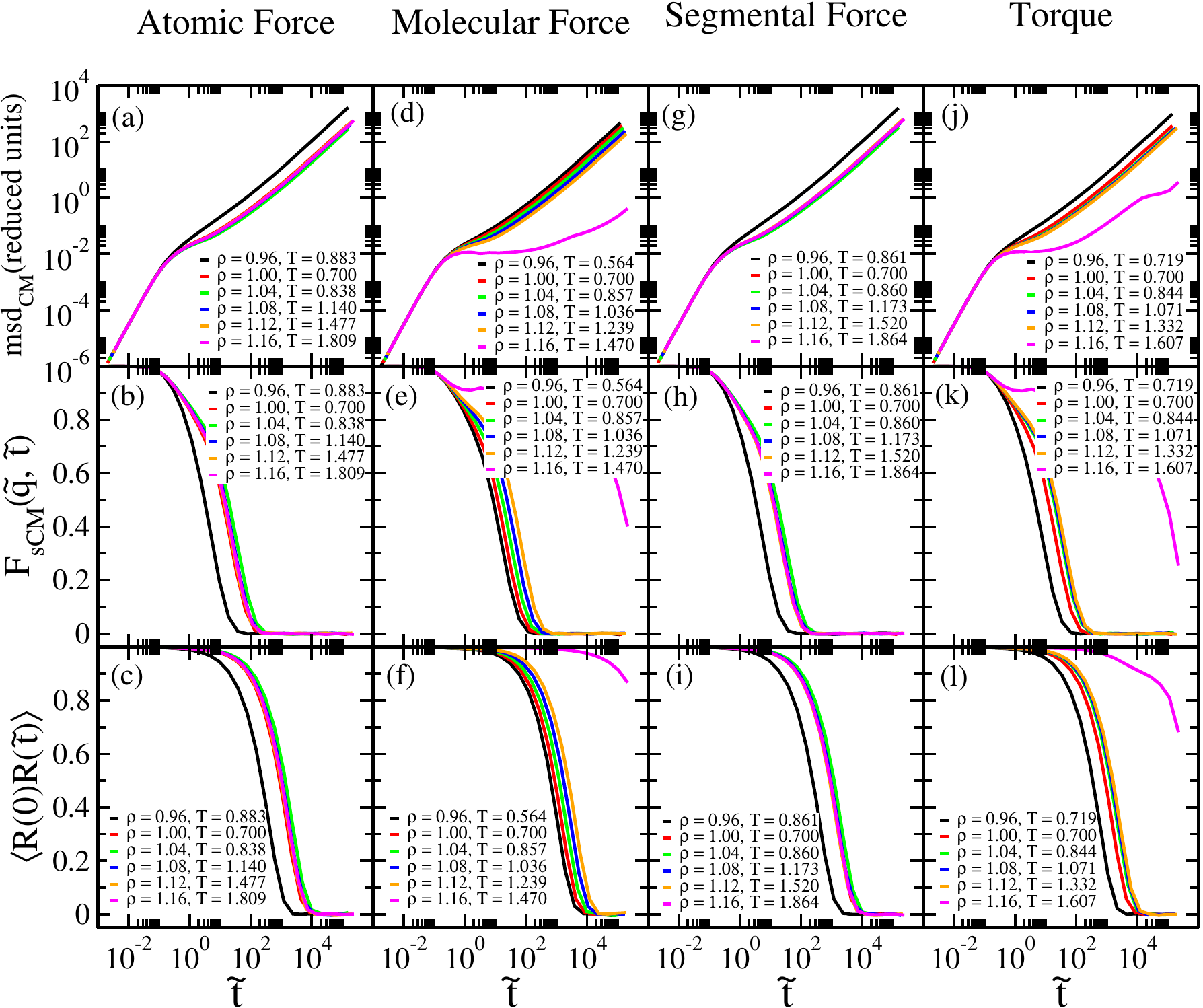}
    \caption{Dynamics of the harmonic-spring LJC model along proposed pseudoisomorphs. The reference state point is $(\rho_1, T_1) = (1.00, 0.700)$ and the density increase is $17\%$. The atomic-force and molecular-force methods both generate good pseudoisomorphs when increasing density as compared to the reference state point. For both methods, decreasing the density leads to a higher temperature, which obviously leads to too fast dynamics. We currently have no explanation as to why this happens. Neither the molecular-force method nor the torque method results in good pseudoisomorphs. As density is increased the predicted temperatures for both methods get smaller and smaller compared to those predicted by the atomic- and segmental- methods, leading to slower dynamics. At the highest density (magenta curves) the dynamics gets very much slower, which might indicate a glass-transition and/or (partial) crystallization. Since neither method works well, we have not investigated this further.
    \label{ch3:FIG:LJC_BQ_AS}}
\end{figure}

Next we turn to the 10-bead harmonic-spring LJC model for which one can use not only the three above methods, but also a fourth one based on invariant reduced segmental forces. These are defined as

\begin{equation}
\mathbf{F}_{Seg,j} \equiv \dfrac{1}{d_j} \mathbf{F}_{j} +  \dfrac{1}{d_{j+1}} \mathbf{F}_{j+1},
\end{equation} 
in which $\mathbf{F}_{Seg,j}$ is the total force on particle $j$, i.e., including also the non-bonded interaction contributions, and $d_j$ and $d_{j+1}$ are the number of bonds that particles $j$ and $j+1$ are involved in. Clearly

\begin{equation}\label{ch3:eq:seg_force}
\mathbf{F}_{Mol} = \sum_{j=1}^{9} \mathbf{F}_{Seg,j}\,.
\end{equation}
The results are quite different from the ASD results. Thus the dynamics are to a good approximation invariant for the atomic- and segmental-force methods (\fig{ch3:FIG:LJC_BQ_AS} (g), (h), (i)), except at the lowest density from reference point $(\rho_1, T_1) = (1.00, 0.700)$ at which the virial becomes negative. On the other hand, neither the molecular force method (d, e, f) nor the torque method (j, k, l) produce good pseudoisomorphs.

\begin{figure}[htbp!]
\centering
    \includegraphics[width=1\textwidth]{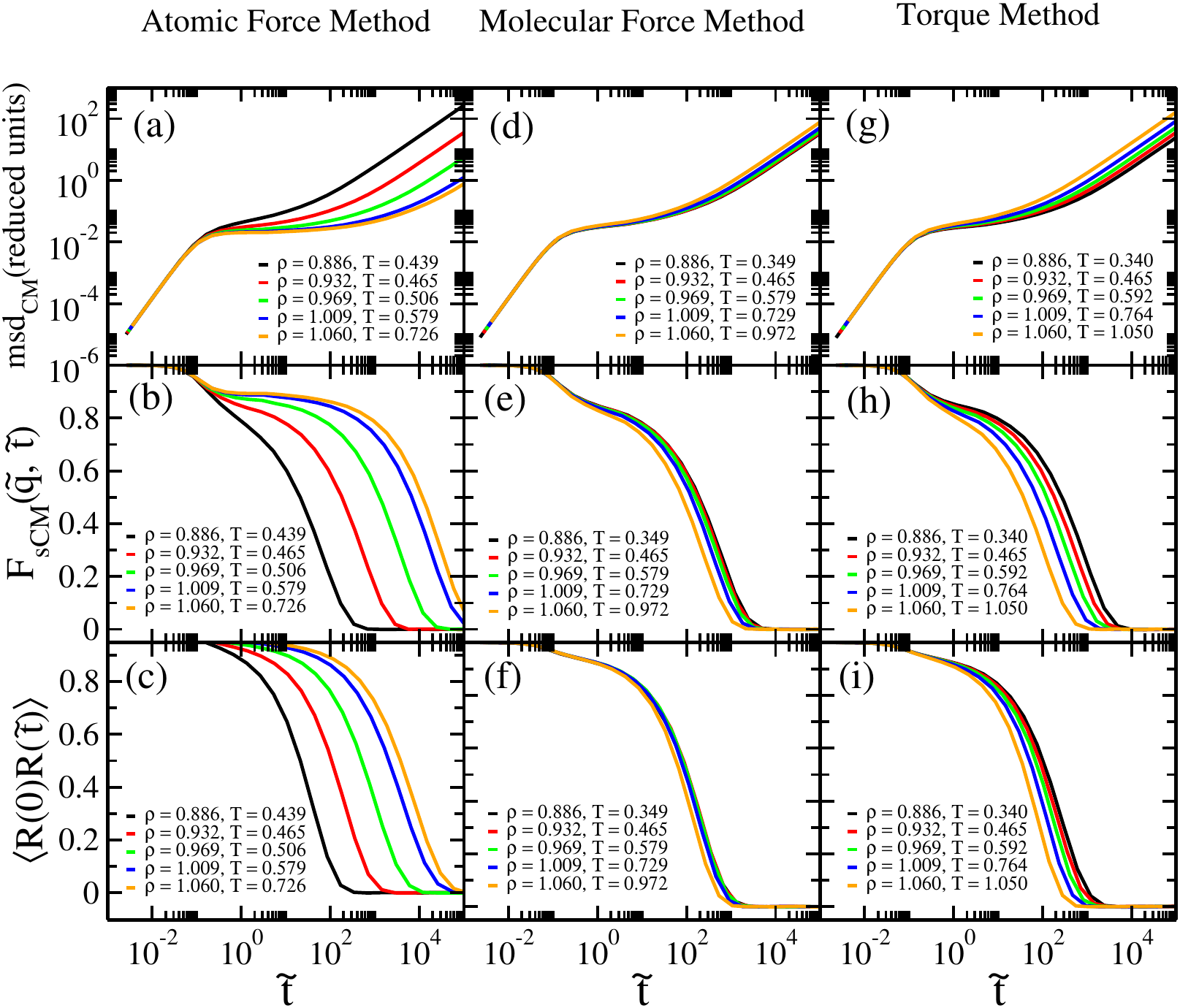}
    \caption{Dynamics of the ASD model along proposed pseudoisomorphs. The densities are here higher than in \fig{ch3:FIG:Dyn_ASD_Low_Den}. The density increase is $19\%$. For all three methods center-of-mass scaling was employed. The reference state point $(\rho_1, T_1) = (0.932, 0.465)$ is taken from Ref. \cite{ols16}. The molecular-force methods works best, but the invariance is not as good as at lower densities (\fig{ch3:FIG:Dyn_ASD_Low_Den}).
    \label{ch3:FIG:ASD_Before_Quench}}
\end{figure}

\section{Pseudoisomorphs in the ASD Model at Higher Densities}

In \fig{ch3:FIG:Dyn_ASD_Low_Den} we applied the three force-based single-configuration methods to the ASD model and found best invariance of the dynamics using the molecular-force and, to a slightly less degree, torque methods. \Fig{ch3:FIG:ASD_Before_Quench} shows the results of applying the three methods at higher densities. The invariance of the dynamics is not as good as at lower densities (\fig{ch3:FIG:Dyn_ASD_Low_Den}). The molecular-force method works best, but no method works as well as the method of Ref. \onlinecite{ols16}, compare \fig{FIG:Pseudo-LJC-ASD}.


\begin{figure}[htbp!]
	\centering
	\resizebox{\textwidth}{!}{
    \begin{tabular}{ll}
        \begin{lpic}[]{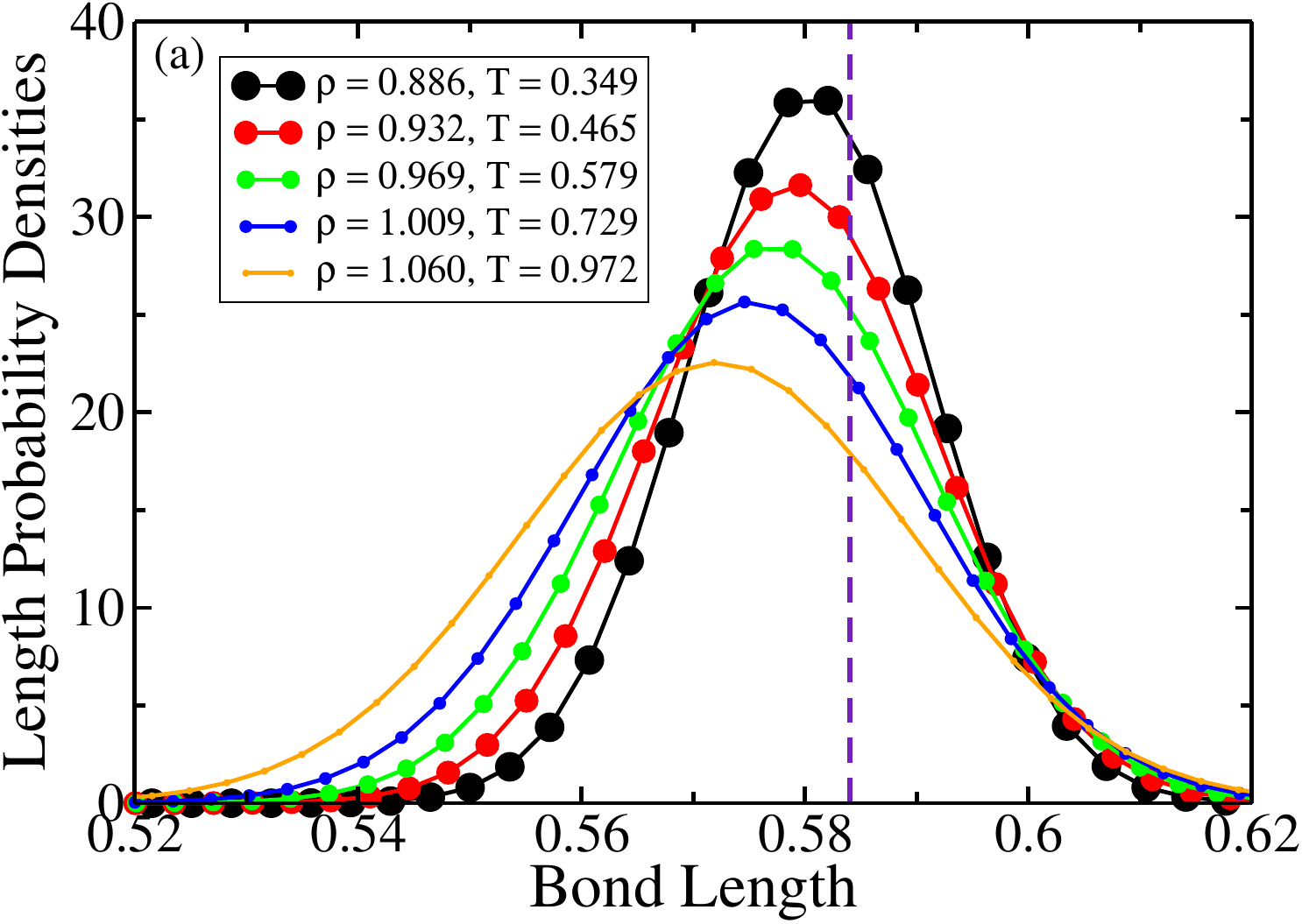}\end{lpic}
        \begin{lpic}[]{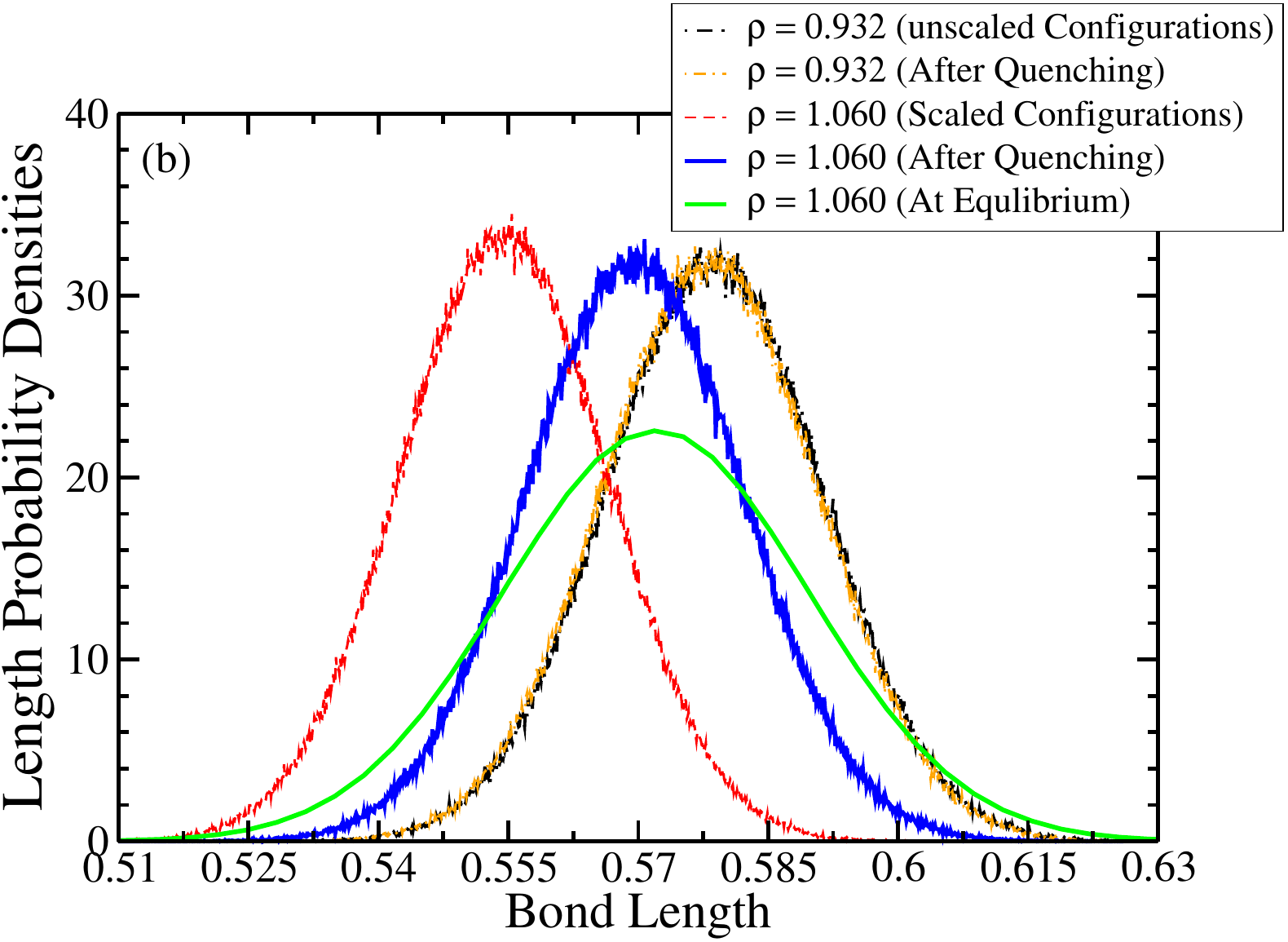}\end{lpic}
    \end{tabular}}
    \caption{(a) Equilibrium distribution of ASD bond lengths at state points identified by the molecular-force method (see \Fig{ch3:FIG:ASD_Before_Quench}(d,e,f)). Bonds are compressed when density is increased. This means that when a configuration is center-of-mass scaled from the reference state point to a (say) higher density, the resulting configuration is not representative of the equilibrium configurations at this higher density.
    (b) Quenching the system according to the constraint conditions (fixed centre-of-mass and orientational direction of all molecules). Black: bond length distribution of original unscaled configuration at the reference state point. Orange: applying the quenching procedure to the unscaled configuration leads to virtually the same distribution. Red:  atomic scaling to the new higher density leads to bond lengths that are considerably smaller than the ones in equilibrium at the new state point (green curve). Blue: center-of-mass scaling to the new higher density followed by the quenching procedure leads to a bond length distribution that is close to the one in equilibrium at the new state point (green curve). Note: the procedure adds the same length 'l' to all bonds, which lead to shift in the distribution leaving the shape invariant.
	\label{ch3:FIG:bond_F_M_ASD}
	}
\end{figure}

\Fig{ch3:FIG:bond_F_M_ASD}(a) shows the distributions of bond lengths for simulations at the state points generated by the molecular-force method. The bonds are compressed when the density is increased, an effect that is not seen at the lower densities of \fig{ch3:FIG:Dyn_ASD_Low_Den}. This means that when a configuration from the reference state point is scaled to a higher density, the scaled configuration is not representative of equilibrium configurations at the new state point because the bonds are too long, an issue that becomes more serious at higher densities.

To eliminate the effects of the harmonic bonds, we introduce the following procedure: For a given configuration fix the center-of-mass and orientation of each molecule. For this ``constrained'' system add a length $l$ to all bond lengths and minimize the potential energy with respect to $l$. We refer to this procedure as ``quenching'', but note that the minimization only involves a single variable, the length $l$ added to all bonds. When applied to an unscaled configuration, the bond length distribution remains largely unchanged (black and orange curves in \fig{ch3:FIG:bond_F_M_ASD} (b)). Now, consider the scaling of the original configuration to higher density, with the aim to achieve a bond length distribution close to that of the equilibrium at the new state point (green curve). Using center-of-mass scaling of the molecules leaves the bond lengths unchanged (black curve), whereas atomic scaling lead to too small bond lengths (red curve). Applying the quenching procedure after center-of-mass scaling shifts the distribution to shorter bond lengths (from the black line to the blue line), and the average bond length approaches that of the equilibrium distribution at the higher density (green line). This effectively eliminates the non-scaling degrees of freedom thought to cause the poor invariance observed in \fig{ch3:FIG:ASD_Before_Quench}.


\begin{figure}[htbp!]
	\centering
	\resizebox{\textwidth}{!}{
    \begin{tabular}{ll}
        \begin{lpic}[]{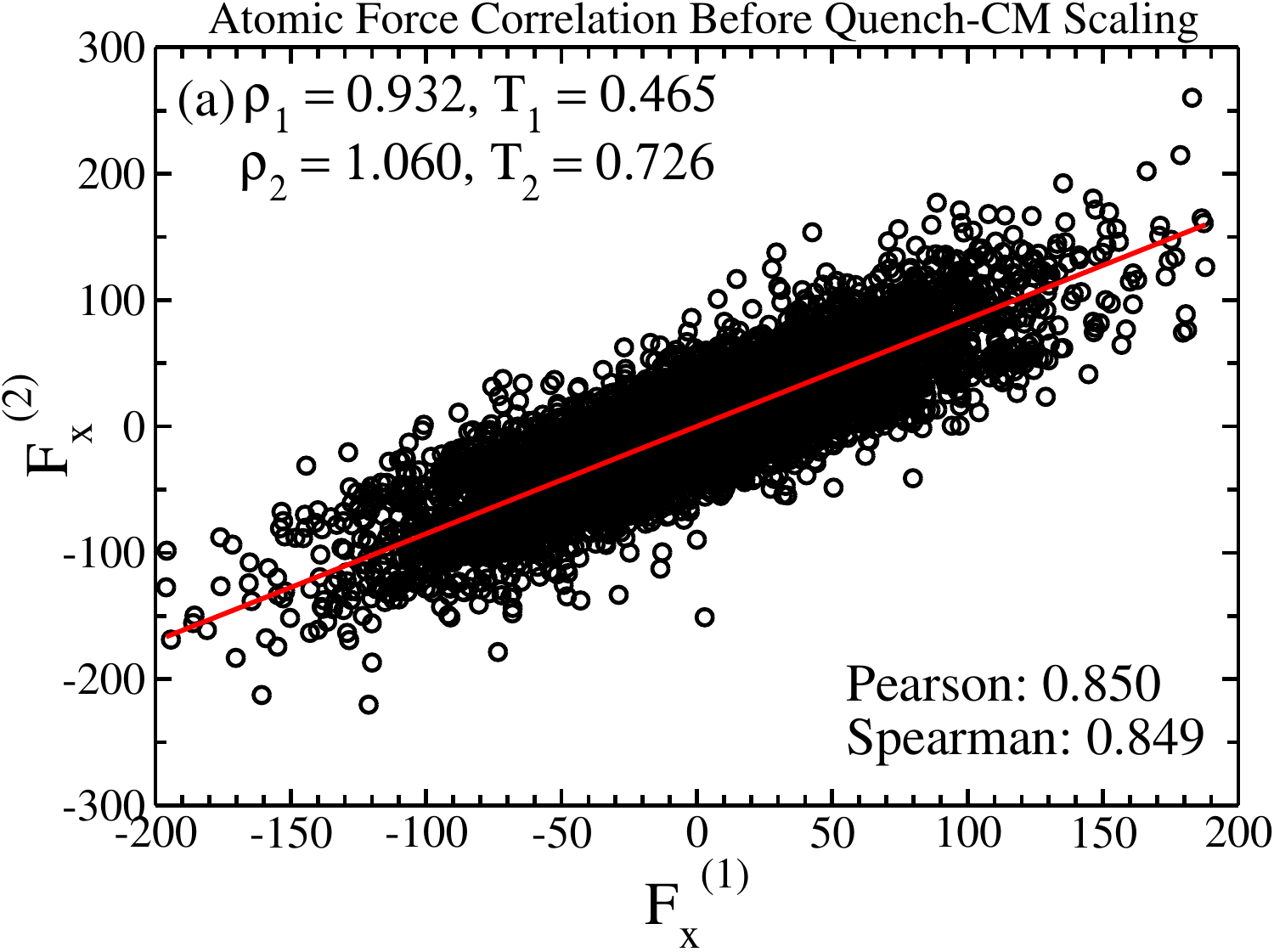}\end{lpic}
        \begin{lpic}[]{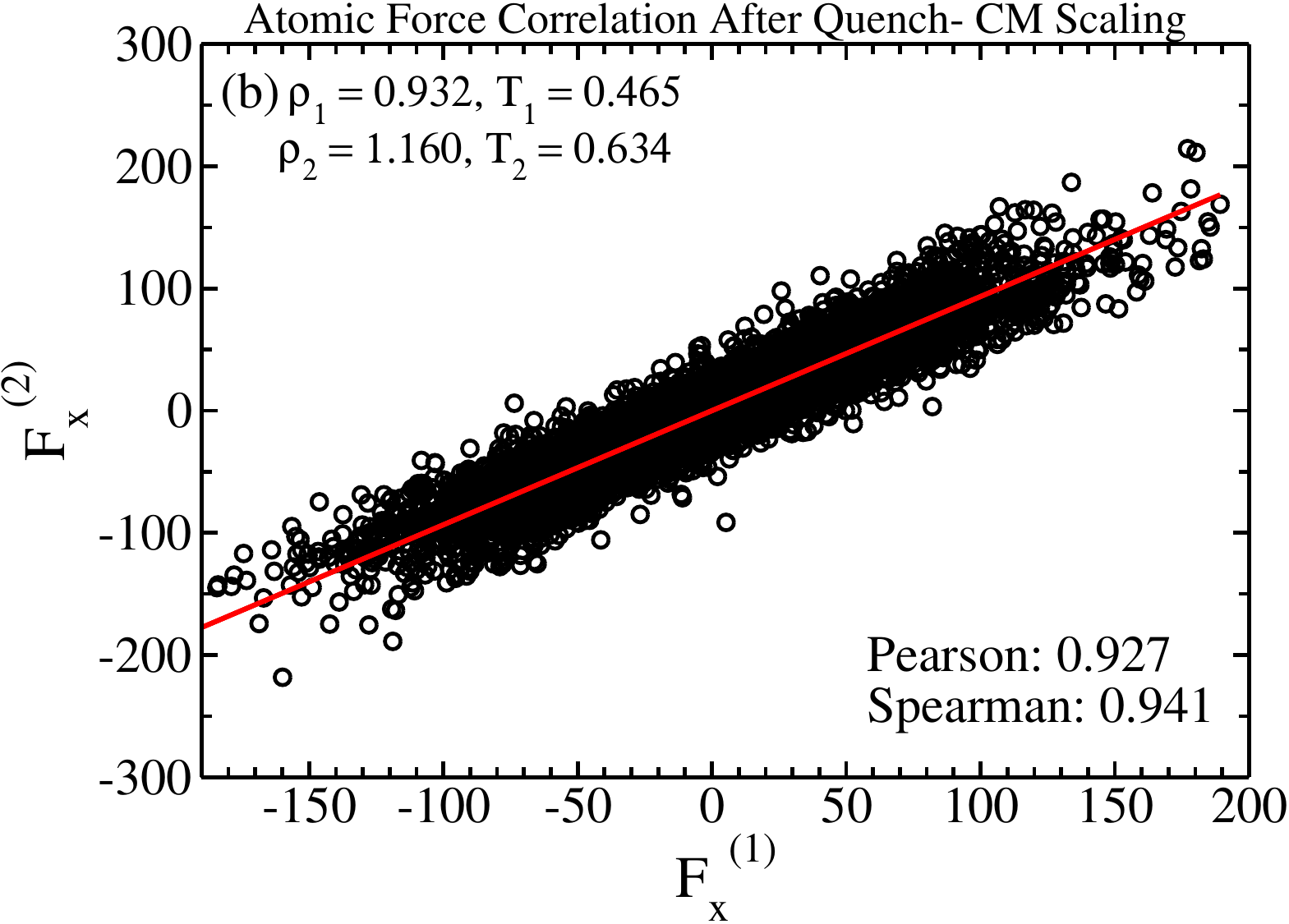}\end{lpic}\\
        \begin{lpic}[]{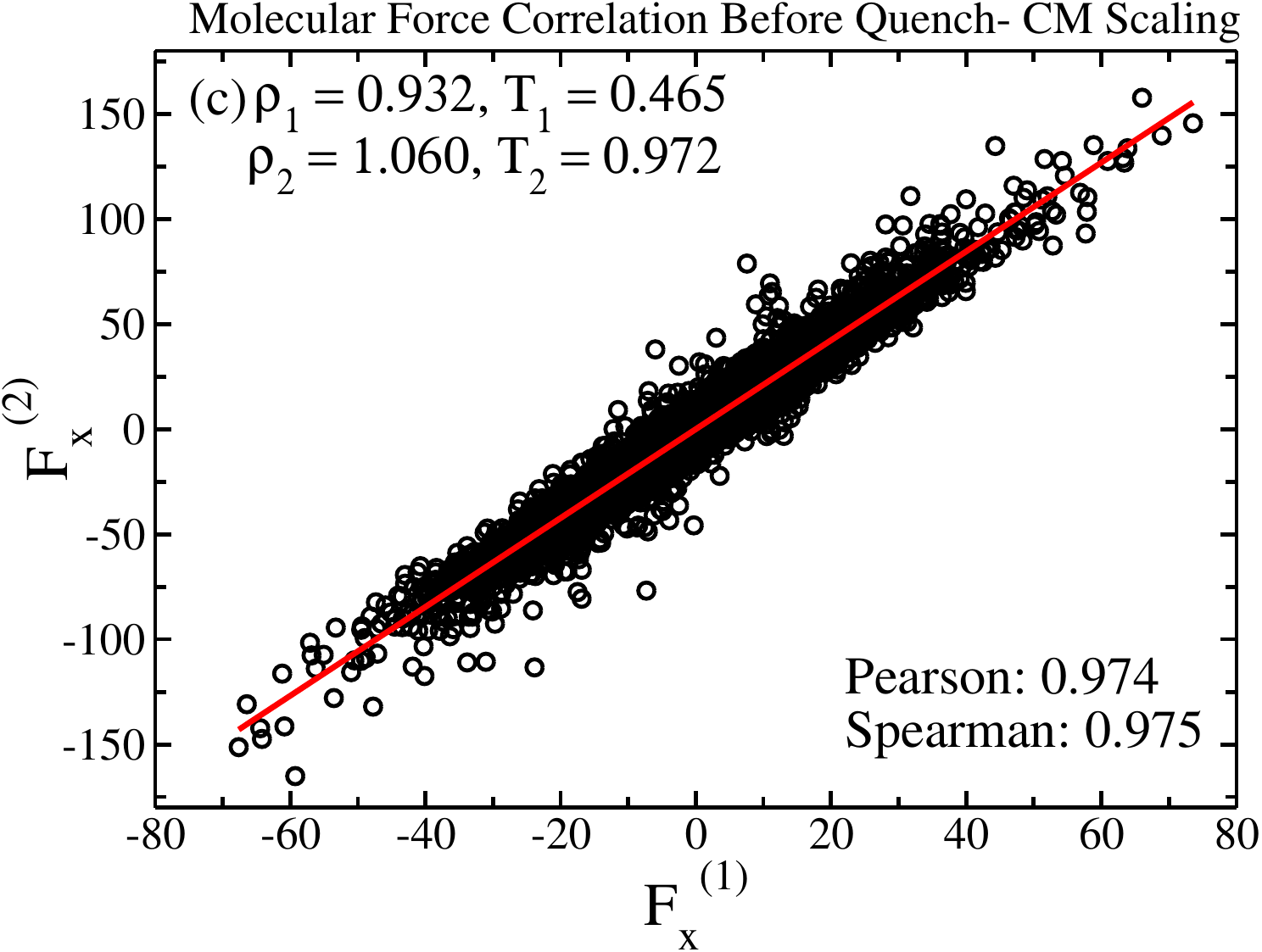}\end{lpic}
        \begin{lpic}[]{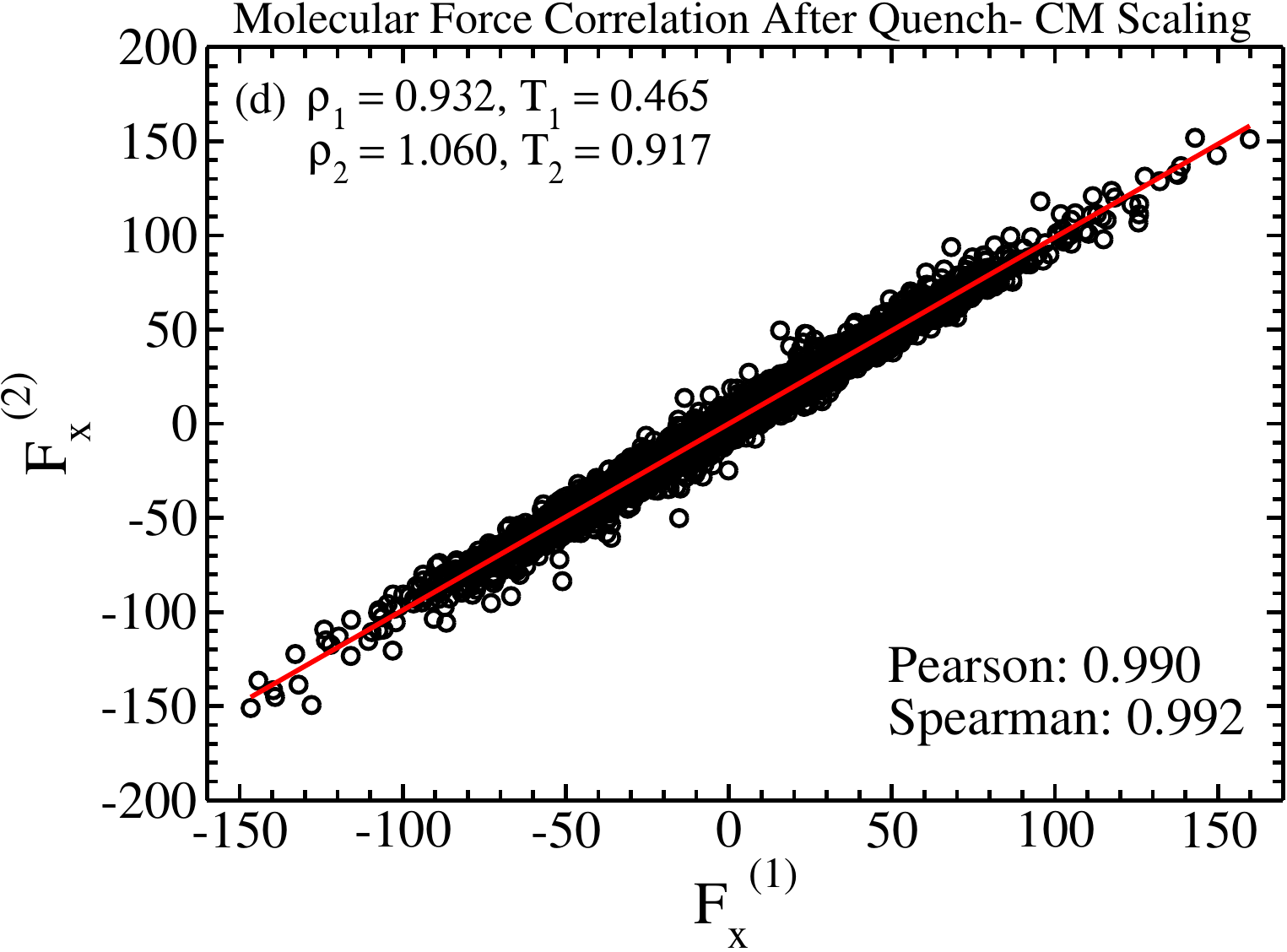}\end{lpic}\\
    \end{tabular}}
	\caption{Correlation of atomic-force components, (a) and (b), and molecular-force components, (c) and (d), for the ASD model. A configuration is taken from a simulation at $(\rho_1, T_1) = (0.932, 0.465)$ and $T_2$ as previously calculated by means of \eq{eq:pre_temp}. For both the atomic-force and molecular-force methods we find significantly better correlations between scaled and non-scaled forces components when applying a quench. Note that only the molecular forces fulfill the criterion that the Pearson and Spearman correlation coefficients are both above 0.95 \cite{sch22}.
	\label{ch3:FIG:Force_Tor_Cor_Spr}
	}
\end{figure}

Based on pairs of quenched configurations, one may again apply the atomic-force, molecular-force, and torque methods to generate state points with, hopefully, same dynamics. Specifically, we proceeded as follows. A single configuration is selected from an equilibrium simulation at the reference state point (density $\rho_1$). This configuration is scaled uniformly to the density of interest, $\rho_2$. Both scaled and unscaled configurations were then quenched as described above in order to eliminate the bond vibrational degrees of freedom. After this the relevant forces / torques were evaluated and $T_2$ determined from \eq{eq:pre_temp} and \eq{eq:pre_tor}, respectively. \Fig{ch3:FIG:Force_Tor_Cor_Spr}(a) shows the forces of a single scaled configuration versus those of the unscaled configuration before quenching, while (b) demonstrates better correlation after quenching; (c) and (d) show the correlations between the  center-of-mass forces before and after quenching. The quench method leads to significantly better correlation, with correlation coefficient increasing from 0.850 to 0.934 in the atomic-force case and from 0.975 to 0.990 in the molecular (center-of-mass) force case.

\begin{figure}[htbp!]
	\centering
	\includegraphics[width=14cm]{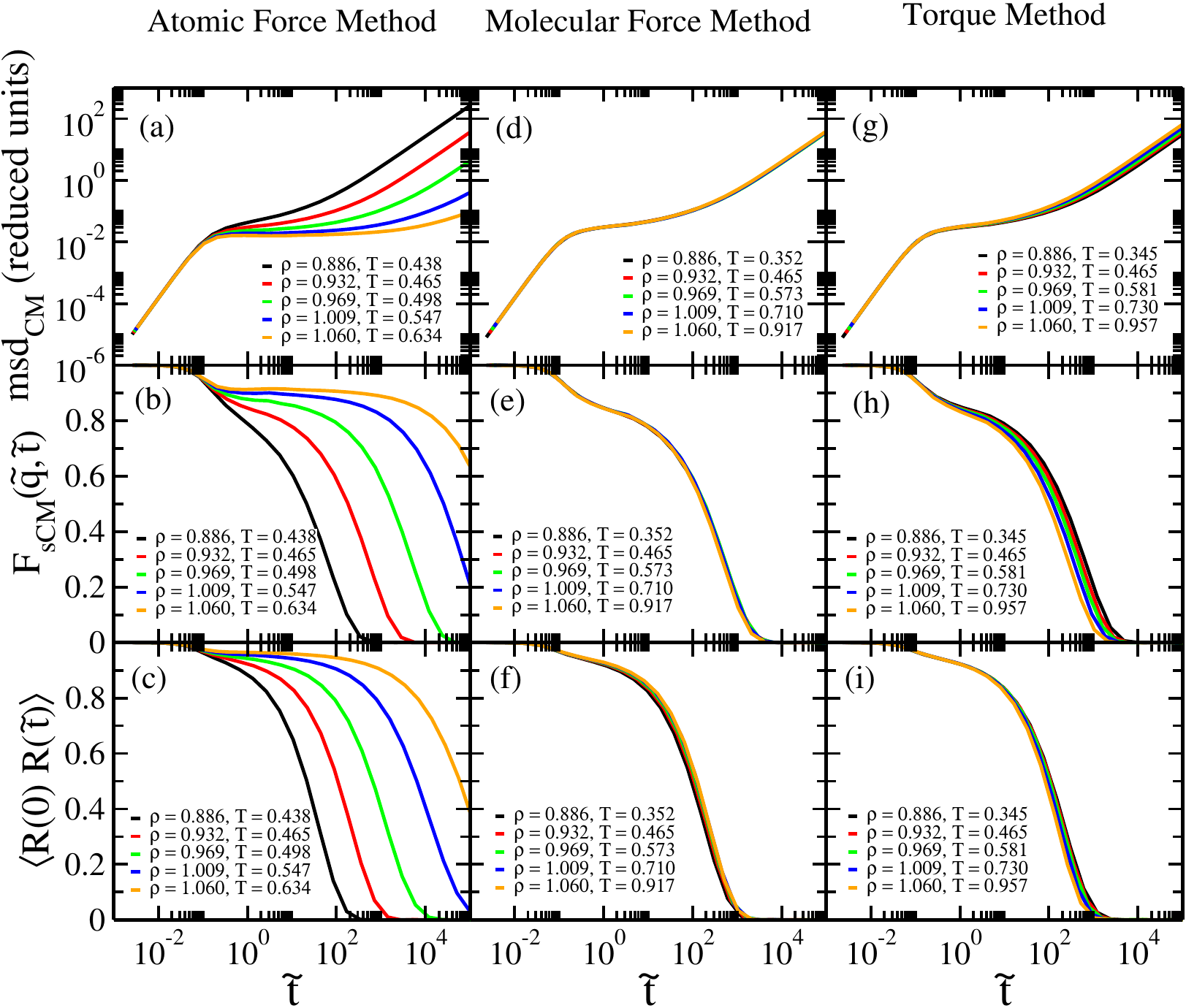}
	\caption{\label{FIG6} Testing the ASD model for invariance of the reduced dynamics using the three quench methods for generating pseudoisomorphs based on a single configuration from $(\rho_1, T_1) = (0.932, 0.465)$. In contrast to \fig{ch3:FIG:ASD_Before_Quench}, the scaled and non-scaled configurations were here quenched to a potential-energy minimum in order to eliminate the harmonic bond degrees of freedom; after which $T_2$ was determined as in \fig{ch3:FIG:ASD_Before_Quench}.
	}
\end{figure}

\begin{figure}[htbp!]
	\centering
	\includegraphics[width=14cm]{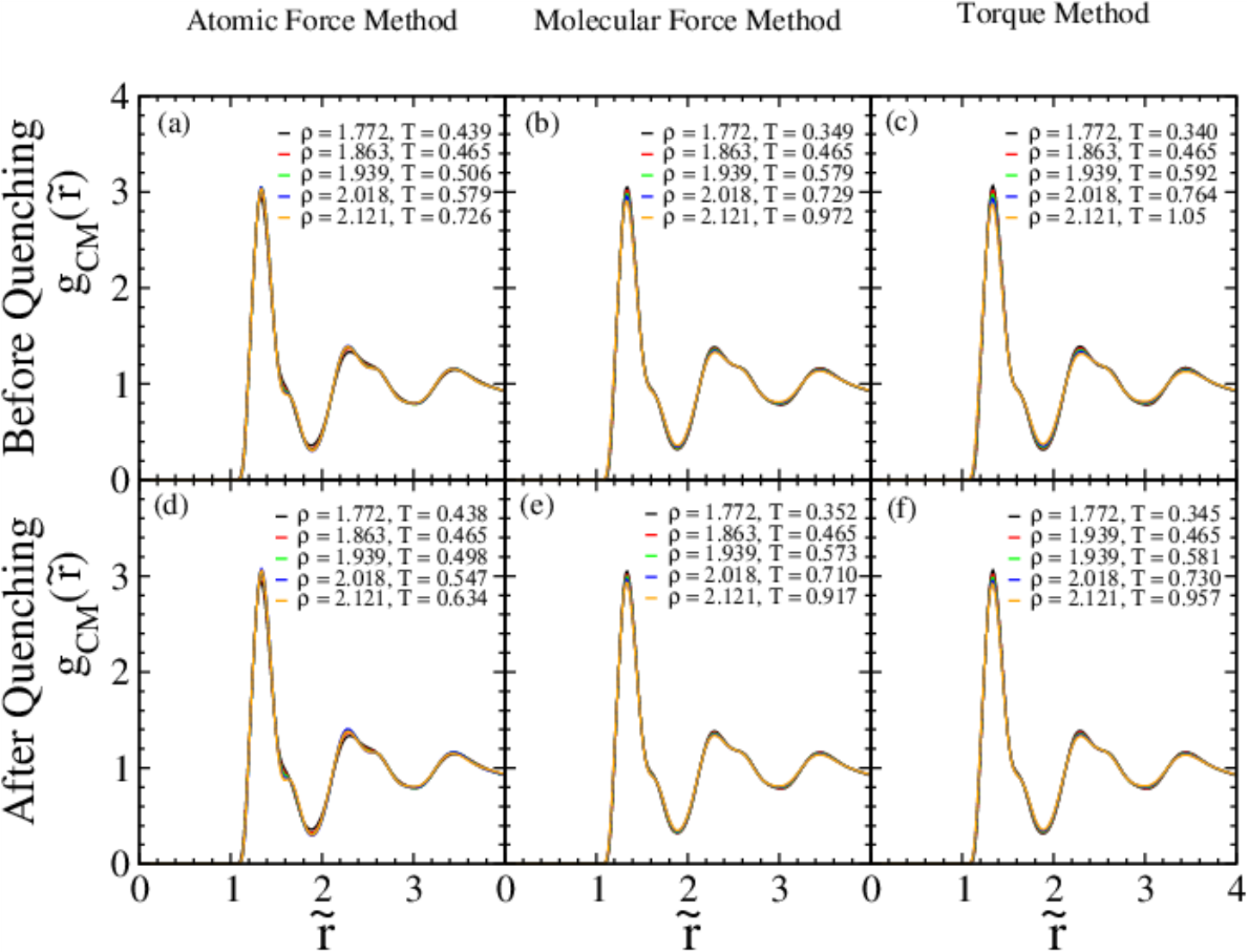}
	\caption{\label{FIG7} Testing the ASD model for invariance of the reduced-unit radial distribution function for the three different methods, without and with quenching, using $(\rho_1, T_1) = (0.932, 0.465)$ as reference state point.  
	(a), (b), (c) show results for state points generated by the atomic-force, molecular-force, and torque methods;
	(d), (e), (f) show results for state points generated by the same methods after minimization.
	In all cases, structure is invariant to a good approximation.
	}
\end{figure}

The effect of the quenching procedure is tested in \fig{FIG6}, which is analogous to \fig{ch3:FIG:ASD_Before_Quench} except that the temperatures $T_2$ are calculated based on quenched configurations. The best results are obtained with the molecular-force method, which was also the one that worked best in \fig{ch3:FIG:Dyn_ASD_Low_Den}. For this method we find excellent collapse of the reduced center-of-mass mean-square displacement as a function of time, as well as of the center-of-mass incoherent intermediate scattering function. The directional time-autocorrelation function shows a slightly worse collapse, but is nevertheless significantly better than without quenching. Comparing the results of the torque method with and without quenching shows that quenching also here significantly improves the invariances. -- Using atomic scaling gives the same results for the molecular force and torque methods after the system has been quenched (Table \ref{ch3:tab:ASD_Spr_AM_AS}), while the atomic-force method gives results different from \fig{FIG6}.

\begin{table}[htbp!]
	\centering
	\caption{Predicted temperature using atomic scaling after quenching (ASD model). Note that the molecular-force and torque methods give the same results as when center-of-mass scaling is used (\fig{FIG6}). $\rho=0.932$ is the reference density in all cases.
	\label{ch3:tab:ASD_Spr_AM_AS}
	}
	\begin{tabular}{cccccc}
		\hline \hline 
		density & \(T(F_{Atomic}) \) & \( T(F_{Mol}) \) & Torque \\
		\hline 
		0.886 & 0.444 & 0.352 & 0.345 \\
		0.932 & 0.465 & 0.465 & 0.465 \\
		0.969 & 0.494 & 0.573 & 0.581 \\
		1.009 & 0.540 & 0.710 & 0.730 \\
		1.060 & 0.624 & 0.917 &  0.957 \\
		\hline 
	\end{tabular}
	\vspace{1ex}
\end{table}

\section{Discussion}

Genuine isomorphs do not exist in systems without strong virial potential-energy correlations, e.g., systems of molecules with harmonic bonds. For the ASD model and the flexible LJC model with harmonic bonds curves nevertheless exist along which the structure and dynamics are invariant to a good approximation, so-called pseudoisomorphs \cite{ols16}. These curves do not have invariant excess entropy, but otherwise behave much like isomorphs by having approximately invariant structure and dynamics in reduced units. It would be interesting to investigate whether some coarse-grained description corresponds to an excess entropy that is invariant along the pseudoisomorphs in this paper, but we have not attempted this and instead prioritized to directly search for methods that identify pseudoisomorphs.

A previously discussed method for tracing out pseudoisomorphs in the thermodynamic phase diagram works well, but is quite complicated to apply in practice\cite{ols16}. The present paper explored the possibility of tracing out pseudoisomorphs based on the simple requirement of invariant length of the reduced force vector of a single configuration after scaling. Although the focus above was on the dynamics, we note that for all three methods discussed the structure is also invariant to a good approximation, both with and without quenching (\fig{FIG7}).

Several closely related but different concepts have been introduced the last couple of decades in order to rationalize important findings of invariant structure and/or dynamics in the thermodynamic phase diagram. The present case of pseudoisomorphs for two realistic molecular models presents a good occasion for summarizing these concepts. One speaks about

\begin{itemize}
	\item \textit{Excess-entropy scaling} when the (reduced-unit) transport coefficients are invariant along the lines of constant excess entropy \cite{ros77,dyr18}. Here DC signals the zero-frequency limit of, e.g., the diffusion coefficient or viscosity;
	\item \textit{Density scaling} when invariant transport coefficients are observed along specific lines in the phase diagram, often described as $\rho^\gamma/T=$Const. \cite{alb04,rol05};
	\item \textit{Isochronal superposition} when entire linear-response relaxation functions are invariant along specific lines in the phase diagram \cite{rol03};
	\item \textit{Isomorph invariance} when structure and dynamics are invariant along the lines of constant excess entropy then termed isomorphs \cite{IV,sch14,dyr18a};
	\item \textit{Pseudoisomorphs} when structure and dynamics are invariant along lines that are not of constant excess entropy \cite{vel14};
	\item \textit{Isodynes} when dynamics, but not structure, is invariant along specific lines \cite{knu21,knu24}.
\end{itemize}
The most recent additions -- pseudoisomorphs and isodynes -- were introduced to describe liquids of more complex molecules, often with internal degrees of freedom. The vision is that some modification of isomorph theory, e.g., arrived at via a suitable coarse-graining focusing on the most important degrees of freedom, may apply also here. At this point in time, however, it is clear that more work is needed in terms of gathering data for many different models to arrive at a coherent picture. Even the relation between pseudoisomorphs and isodynes is not clear. For instance, it is an open question whether large molecules with many internal degrees of freedom are pseudoisomorphs or just isodynes, e.g., without structural invariants. This will depend on the relevant coarse-graining; in fact, systems previously classified as isodynes do have important parts of their structure factor being invariant along the same lines as the dynamics \cite{knu21,knu24}.

For the harmonic-bond ASD model the best method for tracing out pseudoisomorphs is the molecular-force method with center-of-mass scaling and quenching of the harmonic bonds (\fig{FIG6}). At low densities, the quenching can be skipped (\fig{ch3:FIG:Dyn_ASD_Low_Den}). We conjecture that this method will work on all small molecules with harmonic bonds, assuming of course that the systems in question have pseudoisomorphs. The fact that quenching improves the tracing out of pseudoisomorphs is significant by indicating that the internal degrees of freedom of medium sized molecular systems can largely be scaled out.

The molecular-force method does not work well for the 10-bead harmonic-bond LJ-chain model for which the best method is the atomic-force method with atomic scaling (\fig{ch3:FIG:LJC_BQ_AS}). That the atomic force method works best for long chains again is intuitively reasonable as a bead will not be aware that it is interacting with one from another chain or one on the same chain (apart from its nearest neighbors).

From a practical point of view one would like to have a single method that works for all molecular models with pseudoisomorphs, including the two models investigated here. The differences found can be interpreted as reflecting differences regarding which forces are important for the motion of the molecules. Despite the harmonic bonds, the ASD is relatively stiff and its motion is largely governed by the center-of-mass ``molecular'' force and the torque. If the scaling were perfect, both the molecular-force and the torque methods would lead to invariant dynamics. The fact that we find the molecular-force method to work better than the torque method can be interpreted as the molecular forces being more important than the torques for the motion of the molecules. It would be interesting to investigate whether molecular models exist for which the torque method works better than the molecular-force method. The molecular-force method does not work well for the LJC model. We interpret this as reflecting the fact that the center-of-mass motion does not determine the motion of this complex, flexible molecule, which should rather be be thought of in terms of the motion of its beads.

In summary, simple single-configuration force-based methods exist for tracing out pseudoisomorphs, which work as well as the complicated method of diagonalizing the Hessian before and after scaling.\cite{ols16} No single-configuration method of general validity has been identified, however. More work is needed to investigate these and other methods. In this regard, we find it encouraging that the introduction of quenches after scaling improves all the methods studied.

\acknowledgments{This work was supported by the VILLUM Foundation's \textit{Matter} grant (VIL16515).}

\section*{REFERENCES}


\begin{thebibliography}{49}%
	\makeatletter
	\providecommand \@ifxundefined [1]{%
		\@ifx{#1\undefined}
	}%
	\providecommand \@ifnum [1]{%
		\ifnum #1\expandafter \@firstoftwo
		\else \expandafter \@secondoftwo
		\fi
	}%
	\providecommand \@ifx [1]{%
		\ifx #1\expandafter \@firstoftwo
		\else \expandafter \@secondoftwo
		\fi
	}%
	\providecommand \natexlab [1]{#1}%
	\providecommand \enquote  [1]{``#1''}%
	\providecommand \bibnamefont  [1]{#1}%
	\providecommand \bibfnamefont [1]{#1}%
	\providecommand \citenamefont [1]{#1}%
	\providecommand \href@noop [0]{\@secondoftwo}%
	\providecommand \href [0]{\begingroup \@sanitize@url \@href}%
	\providecommand \@href[1]{\@@startlink{#1}\@@href}%
	\providecommand \@@href[1]{\endgroup#1\@@endlink}%
	\providecommand \@sanitize@url [0]{\catcode `\\12\catcode `\$12\catcode
		`\&12\catcode `\#12\catcode `\^12\catcode `\_12\catcode `\%12\relax}%
	\providecommand \@@startlink[1]{}%
	\providecommand \@@endlink[0]{}%
	\providecommand \url  [0]{\begingroup\@sanitize@url \@url }%
	\providecommand \@url [1]{\endgroup\@href {#1}{\urlprefix }}%
	\providecommand \urlprefix  [0]{URL }%
	\providecommand \Eprint [0]{\href }%
	\providecommand \doibase [0]{http://dx.doi.org/}%
	\providecommand \selectlanguage [0]{\@gobble}%
	\providecommand \bibinfo  [0]{\@secondoftwo}%
	\providecommand \bibfield  [0]{\@secondoftwo}%
	\providecommand \translation [1]{[#1]}%
	\providecommand \BibitemOpen [0]{}%
	\providecommand \bibitemStop [0]{}%
	\providecommand \bibitemNoStop [0]{.\EOS\space}%
	\providecommand \EOS [0]{\spacefactor3000\relax}%
	\providecommand \BibitemShut  [1]{\csname bibitem#1\endcsname}%
	\let\auto@bib@innerbib\@empty
	\bibitem [{\citenamefont {Gnan}\ \emph {et~al.}(2009)\citenamefont {Gnan},
		\citenamefont {Schr{\o}der}, \citenamefont {Pedersen}, \citenamefont
		{Bailey},\ and\ \citenamefont {Dyre}}]{IV}%
	\BibitemOpen
	\bibfield  {author} {\bibinfo {author} {\bibfnamefont {N.}~\bibnamefont
			{Gnan}}, \bibinfo {author} {\bibfnamefont {T.~B.}\ \bibnamefont
			{Schr{\o}der}}, \bibinfo {author} {\bibfnamefont {U.~R.}\ \bibnamefont
			{Pedersen}}, \bibinfo {author} {\bibfnamefont {N.~P.}\ \bibnamefont
			{Bailey}}, \ and\ \bibinfo {author} {\bibfnamefont {J.~C.}\ \bibnamefont
			{Dyre}},\ }\bibfield  {title} {\enquote {\bibinfo {title} {Pressure-energy
				correlations in liquids. {IV. ``Isomorphs''} in liquid phase diagrams},}\
	}\href {\doibase 10.1063/1.3265957} {\bibfield  {journal} {\bibinfo
			{journal} {J. Chem. Phys.}\ }\textbf {\bibinfo {volume} {131}},\ \bibinfo
		{pages} {234504} (\bibinfo {year} {2009})}\BibitemShut {NoStop}%
	\bibitem [{\citenamefont {Hansen}\ and\ \citenamefont
		{McDonald}(2013)}]{han13}%
	\BibitemOpen
	\bibfield  {author} {\bibinfo {author} {\bibfnamefont {J.-P.}\ \bibnamefont
			{Hansen}}\ and\ \bibinfo {author} {\bibfnamefont {I.~R.}\ \bibnamefont
			{McDonald}},\ }\href@noop {} {\emph {\bibinfo {title} {{Theory of Simple
					Liquids: With Applications to Soft Matter}}}},\ \bibinfo {edition} {4th}\
	ed.\ (\bibinfo  {publisher} {Academic, New York},\ \bibinfo {year}
	{2013})\BibitemShut {NoStop}%
	\bibitem [{\citenamefont {Malins}\ \emph {et~al.}(2013)\citenamefont {Malins},
		\citenamefont {Eggers},\ and\ \citenamefont {Royall}}]{mal13}%
	\BibitemOpen
	\bibfield  {author} {\bibinfo {author} {\bibfnamefont {A.}~\bibnamefont
			{Malins}}, \bibinfo {author} {\bibfnamefont {J.}~\bibnamefont {Eggers}}, \
		and\ \bibinfo {author} {\bibfnamefont {C.~P.}\ \bibnamefont {Royall}},\
	}\bibfield  {title} {\enquote {\bibinfo {title} {Investigating isomorphs with
				the topological cluster classification},}\ }\href@noop {} {\bibfield
		{journal} {\bibinfo  {journal} {J. Chem. Phys.}\ }\textbf {\bibinfo {volume}
			{139}},\ \bibinfo {pages} {234505} (\bibinfo {year} {2013})}\BibitemShut
	{NoStop}%
	\bibitem [{\citenamefont {Flenner}\ \emph {et~al.}(2014)\citenamefont
		{Flenner}, \citenamefont {Staley},\ and\ \citenamefont {Szamel}}]{fle14}%
	\BibitemOpen
	\bibfield  {author} {\bibinfo {author} {\bibfnamefont {E.}~\bibnamefont
			{Flenner}}, \bibinfo {author} {\bibfnamefont {H.}~\bibnamefont {Staley}}, \
		and\ \bibinfo {author} {\bibfnamefont {G.}~\bibnamefont {Szamel}},\
	}\bibfield  {title} {\enquote {\bibinfo {title} {{Universal Features of
					Dynamic Heterogeneity in Supercooled Liquids}},}\ }\href@noop {} {\bibfield
		{journal} {\bibinfo  {journal} {Phys. Rev. Lett.}\ }\textbf {\bibinfo
			{volume} {112}},\ \bibinfo {pages} {097801} (\bibinfo {year}
		{2014})}\BibitemShut {NoStop}%
	\bibitem [{\citenamefont {Prasad}\ and\ \citenamefont
		{Chakravarty}(2014)}]{pra14}%
	\BibitemOpen
	\bibfield  {author} {\bibinfo {author} {\bibfnamefont {S.}~\bibnamefont
			{Prasad}}\ and\ \bibinfo {author} {\bibfnamefont {C.}~\bibnamefont
			{Chakravarty}},\ }\bibfield  {title} {\enquote {\bibinfo {title} {Onset of
				simple liquid behaviour in modified water models},}\ }\href {\doibase
		10.1063/1.4870823} {\bibfield  {journal} {\bibinfo  {journal} {J. Chem.
				Phys.}\ }\textbf {\bibinfo {volume} {140}},\ \bibinfo {pages} {164501}
		(\bibinfo {year} {2014})}\BibitemShut {NoStop}%
	\bibitem [{\citenamefont {Schr{\o}der}\ and\ \citenamefont
		{Dyre}(2014)}]{sch14}%
	\BibitemOpen
	\bibfield  {author} {\bibinfo {author} {\bibfnamefont {T.~B.}\ \bibnamefont
			{Schr{\o}der}}\ and\ \bibinfo {author} {\bibfnamefont {J.~C.}\ \bibnamefont
			{Dyre}},\ }\bibfield  {title} {\enquote {\bibinfo {title} {Simplicity of
				condensed matter at its core: Generic definition of a {Roskilde}-simple
				system},}\ }\href {\doibase http://dx.doi.org/10.1063/1.4901215} {\bibfield
		{journal} {\bibinfo  {journal} {J. Chem. Phys.}\ }\textbf {\bibinfo {volume}
			{141}},\ \bibinfo {pages} {204502} (\bibinfo {year} {2014})}\BibitemShut
	{NoStop}%
	\bibitem [{\citenamefont {Heyes}\ \emph {et~al.}(2015)\citenamefont {Heyes},
		\citenamefont {Dini},\ and\ \citenamefont {Branka}}]{hey15}%
	\BibitemOpen
	\bibfield  {author} {\bibinfo {author} {\bibfnamefont {D.~M.}\ \bibnamefont
			{Heyes}}, \bibinfo {author} {\bibfnamefont {D.}~\bibnamefont {Dini}}, \ and\
		\bibinfo {author} {\bibfnamefont {A.~C.}\ \bibnamefont {Branka}},\ }\bibfield
	{title} {\enquote {\bibinfo {title} {Scaling of {Lennard-Jones} liquid
				elastic moduli, viscoelasticity and other properties along fluid-solid
				coexistence},}\ }\href {\doibase 10.1002/pssb.201451695} {\bibfield
		{journal} {\bibinfo  {journal} {Phys. Status Solidi (b)}\ }\textbf {\bibinfo
			{volume} {252}},\ \bibinfo {pages} {1514--1525} (\bibinfo {year}
		{2015})}\BibitemShut {NoStop}%
	\bibitem [{\citenamefont {Khrapak}\ \emph {et~al.}(2016)\citenamefont
		{Khrapak}, \citenamefont {Klumov}, \citenamefont {Couedel},\ and\
		\citenamefont {Thomas}}]{khr16}%
	\BibitemOpen
	\bibfield  {author} {\bibinfo {author} {\bibfnamefont {S.~A.}\ \bibnamefont
			{Khrapak}}, \bibinfo {author} {\bibfnamefont {B.}~\bibnamefont {Klumov}},
		\bibinfo {author} {\bibfnamefont {L.}~\bibnamefont {Couedel}}, \ and\
		\bibinfo {author} {\bibfnamefont {H.~M.}\ \bibnamefont {Thomas}},\ }\bibfield
	{title} {\enquote {\bibinfo {title} {On the long-waves dispersion in
				{Yukawa} systems},}\ }\href@noop {} {\bibfield  {journal} {\bibinfo
			{journal} {Phys. Plasmas}\ }\textbf {\bibinfo {volume} {23}},\ \bibinfo {eid}
		{023702} (\bibinfo {year} {2016})}\BibitemShut {NoStop}%
	\bibitem [{\citenamefont {Kaskosz}\ \emph {et~al.}(2023)\citenamefont
		{Kaskosz}, \citenamefont {Koperwas}, \citenamefont {Grzybowski},\ and\
		\citenamefont {Paluch}}]{kas23}%
	\BibitemOpen
	\bibfield  {author} {\bibinfo {author} {\bibfnamefont {F.}~\bibnamefont
			{Kaskosz}}, \bibinfo {author} {\bibfnamefont {K}~\bibnamefont {Koperwas}},
		\bibinfo {author} {\bibfnamefont {A.}~\bibnamefont {Grzybowski}}, \ and\
		\bibinfo {author} {\bibfnamefont {M.}~\bibnamefont {Paluch}},\ }\bibfield
	{title} {\enquote {\bibinfo {title} {The origin of the density scaling
				exponent for polyatomic molecules and the estimation of its value from the
				liquid structure},}\ }\href {\doibase 10.1063/5.0141975} {\bibfield
		{journal} {\bibinfo  {journal} {J. Chem. Phys.}\ }\textbf {\bibinfo {volume}
			{158}},\ \bibinfo {pages} {144503} (\bibinfo {year} {2023})}\BibitemShut
	{NoStop}%
	\bibitem [{\citenamefont {Ingebrigtsen}\ \emph {et~al.}(2012)\citenamefont
		{Ingebrigtsen}, \citenamefont {Schr{\o}der},\ and\ \citenamefont
		{Dyre}}]{ing12b}%
	\BibitemOpen
	\bibfield  {author} {\bibinfo {author} {\bibfnamefont {T.~S.}\ \bibnamefont
			{Ingebrigtsen}}, \bibinfo {author} {\bibfnamefont {T.~B.}\ \bibnamefont
			{Schr{\o}der}}, \ and\ \bibinfo {author} {\bibfnamefont {J.~C.}\ \bibnamefont
			{Dyre}},\ }\bibfield  {title} {\enquote {\bibinfo {title} {Isomorphs in model
				molecular liquids},}\ }\href {\doibase 10.1021/jp2077402} {\bibfield
		{journal} {\bibinfo  {journal} {J. Phys. Chem. B}\ }\textbf {\bibinfo
			{volume} {116}},\ \bibinfo {pages} {1018--1034} (\bibinfo {year}
		{2012})}\BibitemShut {NoStop}%
	\bibitem [{\citenamefont {Attia}\ \emph {et~al.}(2021)\citenamefont {Attia},
		\citenamefont {Dyre},\ and\ \citenamefont {Pedersen}}]{att21}%
	\BibitemOpen
	\bibfield  {author} {\bibinfo {author} {\bibfnamefont {E.}~\bibnamefont
			{Attia}}, \bibinfo {author} {\bibfnamefont {J.~C.}\ \bibnamefont {Dyre}}, \
		and\ \bibinfo {author} {\bibfnamefont {U.~R.}\ \bibnamefont {Pedersen}},\
	}\bibfield  {title} {\enquote {\bibinfo {title} {Extreme case of density
				scaling: {The Weeks-Chandler-Andersen} system at low temperatures},}\ }\href
	{\doibase 10.1103/PhysRevE.103.062140} {\bibfield  {journal} {\bibinfo
			{journal} {Phys. Rev. E}\ }\textbf {\bibinfo {volume} {103}},\ \bibinfo
		{pages} {062140} (\bibinfo {year} {2021})}\BibitemShut {NoStop}%
	\bibitem [{\citenamefont {B{\o}hling}\ \emph {et~al.}(2012)\citenamefont
		{B{\o}hling}, \citenamefont {Ingebrigtsen}, \citenamefont {Grzybowski},
		\citenamefont {Paluch}, \citenamefont {Dyre},\ and\ \citenamefont
		{Schr{\o}der}}]{boh12}%
	\BibitemOpen
	\bibfield  {author} {\bibinfo {author} {\bibfnamefont {L.}~\bibnamefont
			{B{\o}hling}}, \bibinfo {author} {\bibfnamefont {T.~S.}\ \bibnamefont
			{Ingebrigtsen}}, \bibinfo {author} {\bibfnamefont {A.}~\bibnamefont
			{Grzybowski}}, \bibinfo {author} {\bibfnamefont {M.}~\bibnamefont {Paluch}},
		\bibinfo {author} {\bibfnamefont {J.~C.}\ \bibnamefont {Dyre}}, \ and\
		\bibinfo {author} {\bibfnamefont {T.~B.}\ \bibnamefont {Schr{\o}der}},\
	}\bibfield  {title} {\enquote {\bibinfo {title} {Scaling of viscous dynamics
				in simple liquids: {Theory,} simulation and experiment},}\ }\href {\doibase
		10.1088/1367-2630/14/11/113035} {\bibfield  {journal} {\bibinfo  {journal}
			{New J. Phys.}\ }\textbf {\bibinfo {volume} {14}},\ \bibinfo {pages} {113035}
		(\bibinfo {year} {2012})}\BibitemShut {NoStop}%
	\bibitem [{\citenamefont {Bacher}\ \emph {et~al.}(2018)\citenamefont {Bacher},
		\citenamefont {Schr{\o}der},\ and\ \citenamefont {Dyre}}]{EXPII}%
	\BibitemOpen
	\bibfield  {author} {\bibinfo {author} {\bibfnamefont {A.~K.}\ \bibnamefont
			{Bacher}}, \bibinfo {author} {\bibfnamefont {T.~B.}\ \bibnamefont
			{Schr{\o}der}}, \ and\ \bibinfo {author} {\bibfnamefont {J.~C.}\ \bibnamefont
			{Dyre}},\ }\bibfield  {title} {\enquote {\bibinfo {title} {The {EXP}
				pair-potential system. {II.} {Fluid} phase isomorphs},}\ }\href {\doibase
		10.1063/1.5043548} {\bibfield  {journal} {\bibinfo  {journal} {J. Chem.
				Phys.}\ }\textbf {\bibinfo {volume} {149}},\ \bibinfo {pages} {114502}
		(\bibinfo {year} {2018})}\BibitemShut {NoStop}%
	\bibitem [{\citenamefont {Heyes}\ \emph {et~al.}(2019)\citenamefont {Heyes},
		\citenamefont {Dini}, \citenamefont {Costigliola},\ and\ \citenamefont
		{Dyre}}]{hey19}%
	\BibitemOpen
	\bibfield  {author} {\bibinfo {author} {\bibfnamefont {D.~M.}\ \bibnamefont
			{Heyes}}, \bibinfo {author} {\bibfnamefont {D.}~\bibnamefont {Dini}},
		\bibinfo {author} {\bibfnamefont {L.}~\bibnamefont {Costigliola}}, \ and\
		\bibinfo {author} {\bibfnamefont {J.~C.}\ \bibnamefont {Dyre}},\ }\bibfield
	{title} {\enquote {\bibinfo {title} {Transport coefficients of the
				{Lennard-Jones} fluid close to the freezing line},}\ }\href {\doibase
		10.1063/1.5128707} {\bibfield  {journal} {\bibinfo  {journal} {J. Chem.
				Phys.}\ }\textbf {\bibinfo {volume} {151}},\ \bibinfo {pages} {204502}
		(\bibinfo {year} {2019})}\BibitemShut {NoStop}%
	\bibitem [{\citenamefont {Castello}\ \emph {et~al.}(2021)\citenamefont
		{Castello}, \citenamefont {Tolias},\ and\ \citenamefont {Dyre}}]{cas21}%
	\BibitemOpen
	\bibfield  {author} {\bibinfo {author} {\bibfnamefont {F.~L.}\ \bibnamefont
			{Castello}}, \bibinfo {author} {\bibfnamefont {P.}~\bibnamefont {Tolias}}, \
		and\ \bibinfo {author} {\bibfnamefont {J.~C.}\ \bibnamefont {Dyre}},\
	}\bibfield  {title} {\enquote {\bibinfo {title} {Testing the isomorph
				invariance of the bridge functions of {Yukawa} one-component plasmas},}\
	}\href {\doibase 10.1063/5.0036226} {\bibfield  {journal} {\bibinfo
			{journal} {J. Chem. Phys.}\ }\textbf {\bibinfo {volume} {154}},\ \bibinfo
		{pages} {034501} (\bibinfo {year} {2021})}\BibitemShut {NoStop}%
	\bibitem [{\citenamefont {Veldhorst}\ \emph {et~al.}(2014)\citenamefont
		{Veldhorst}, \citenamefont {Dyre},\ and\ \citenamefont
		{Schr{\o}der}}]{vel14}%
	\BibitemOpen
	\bibfield  {author} {\bibinfo {author} {\bibfnamefont {A.~A.}\ \bibnamefont
			{Veldhorst}}, \bibinfo {author} {\bibfnamefont {J.~C.}\ \bibnamefont {Dyre}},
		\ and\ \bibinfo {author} {\bibfnamefont {T.~B.}\ \bibnamefont
			{Schr{\o}der}},\ }\bibfield  {title} {\enquote {\bibinfo {title} {Scaling of
				the dynamics of flexible {Lennard-Jones} chains},}\ }\href {\doibase
		10.1063/1.4888564} {\bibfield  {journal} {\bibinfo  {journal} {J. Chem.
				Phys.}\ }\textbf {\bibinfo {volume} {141}},\ \bibinfo {pages} {054904}
		(\bibinfo {year} {2014})}\BibitemShut {NoStop}%
	\bibitem [{\citenamefont {Xiao}\ \emph {et~al.}(2015)\citenamefont {Xiao},
		\citenamefont {Tofteskov}, \citenamefont {Christensen}, \citenamefont
		{Dyre},\ and\ \citenamefont {Niss}}]{xia15}%
	\BibitemOpen
	\bibfield  {author} {\bibinfo {author} {\bibfnamefont {W.}~\bibnamefont
			{Xiao}}, \bibinfo {author} {\bibfnamefont {J.}~\bibnamefont {Tofteskov}},
		\bibinfo {author} {\bibfnamefont {T.~V.}\ \bibnamefont {Christensen}},
		\bibinfo {author} {\bibfnamefont {J.~C.}\ \bibnamefont {Dyre}}, \ and\
		\bibinfo {author} {\bibfnamefont {K.}~\bibnamefont {Niss}},\ }\bibfield
	{title} {\enquote {\bibinfo {title} {Isomorph theory prediction for the
				dielectric loss variation along an isochrone},}\ }\href@noop {} {\bibfield
		{journal} {\bibinfo  {journal} {J. Non-Cryst. Solids}\ }\textbf {\bibinfo
			{volume} {407}},\ \bibinfo {pages} {190--195} (\bibinfo {year}
		{2015})}\BibitemShut {NoStop}%
	\bibitem [{\citenamefont {Hansen}\ \emph {et~al.}(2018)\citenamefont {Hansen},
		\citenamefont {Sanz}, \citenamefont {Adrjanowicz}, \citenamefont {Frick},\
		and\ \citenamefont {Niss}}]{han18}%
	\BibitemOpen
	\bibfield  {author} {\bibinfo {author} {\bibfnamefont {H.~W.}\ \bibnamefont
			{Hansen}}, \bibinfo {author} {\bibfnamefont {A.}~\bibnamefont {Sanz}},
		\bibinfo {author} {\bibfnamefont {K.}~\bibnamefont {Adrjanowicz}}, \bibinfo
		{author} {\bibfnamefont {B.}~\bibnamefont {Frick}}, \ and\ \bibinfo {author}
		{\bibfnamefont {K.}~\bibnamefont {Niss}},\ }\bibfield  {title} {\enquote
		{\bibinfo {title} {Evidence of a one-dimensional thermodynamic phase diagram
				for simple glass-formers},}\ }\href {\doibase doi:10.1038/s41467-017-02324-3}
	{\bibfield  {journal} {\bibinfo  {journal} {Nat. Commun.}\ }\textbf {\bibinfo
			{volume} {9}},\ \bibinfo {pages} {518} (\bibinfo {year} {2018})}\BibitemShut
	{NoStop}%
	\bibitem [{\citenamefont {Veldhorst}\ \emph {et~al.}(2015)\citenamefont
		{Veldhorst}, \citenamefont {Dyre},\ and\ \citenamefont
		{Schr{\o}der}}]{vel15a}%
	\BibitemOpen
	\bibfield  {author} {\bibinfo {author} {\bibfnamefont {A.~A.}\ \bibnamefont
			{Veldhorst}}, \bibinfo {author} {\bibfnamefont {J.~C.}\ \bibnamefont {Dyre}},
		\ and\ \bibinfo {author} {\bibfnamefont {T.~B.}\ \bibnamefont
			{Schr{\o}der}},\ }\bibfield  {title} {\enquote {\bibinfo {title} {Scaling of
				the dynamics of flexible {Lennard-Jones} chains: {Effects} of harmonic
				bonds},}\ }\href {\doibase 10.1063/1.4934973} {\bibfield  {journal} {\bibinfo
			{journal} {J. Chem. Phys.}\ }\textbf {\bibinfo {volume} {143}},\ \bibinfo
		{pages} {194503} (\bibinfo {year} {2015})}\BibitemShut {NoStop}%
	\bibitem [{\citenamefont {Olsen}\ \emph {et~al.}(2016)\citenamefont {Olsen},
		\citenamefont {Dyre},\ and\ \citenamefont {Schr{\o}der}}]{ols16}%
	\BibitemOpen
	\bibfield  {author} {\bibinfo {author} {\bibfnamefont {A.~E.}\ \bibnamefont
			{Olsen}}, \bibinfo {author} {\bibfnamefont {J.~C.}\ \bibnamefont {Dyre}}, \
		and\ \bibinfo {author} {\bibfnamefont {T.~B.}\ \bibnamefont {Schr{\o}der}},\
	}\bibfield  {title} {\enquote {\bibinfo {title} {Communication:
				{Pseudoisomorphs} in liquids with intramolecular degrees of freedom},}\
	}\href {\doibase 10.1063/1.4972860} {\bibfield  {journal} {\bibinfo
			{journal} {J. Chem. Phys.}\ }\textbf {\bibinfo {volume} {145}},\ \bibinfo
		{pages} {241103} (\bibinfo {year} {2016})}\BibitemShut {NoStop}%
	\bibitem [{\citenamefont {Koperwas}\ and\ \citenamefont
		{Paluch}(2022)}]{kop22}%
	\BibitemOpen
	\bibfield  {author} {\bibinfo {author} {\bibfnamefont {K.}~\bibnamefont
			{Koperwas}}\ and\ \bibinfo {author} {\bibfnamefont {M.}~\bibnamefont
			{Paluch}},\ }\bibfield  {title} {\enquote {\bibinfo {title} {Computational
				evidence for the crucial role of dipole cross-correlations in polar
				glass-forming liquids},}\ }\href {\doibase 10.1103/PhysRevLett.129.025501}
	{\bibfield  {journal} {\bibinfo  {journal} {Phys. Rev. Lett.}\ }\textbf
		{\bibinfo {volume} {129}},\ \bibinfo {pages} {025501} (\bibinfo {year}
		{2022})}\BibitemShut {NoStop}%
	\bibitem [{\citenamefont {Stillinger}\ and\ \citenamefont
		{Weber}(1983)}]{sti83}%
	\BibitemOpen
	\bibfield  {author} {\bibinfo {author} {\bibfnamefont {F.~H.}\ \bibnamefont
			{Stillinger}}\ and\ \bibinfo {author} {\bibfnamefont {T.~A.}\ \bibnamefont
			{Weber}},\ }\bibfield  {title} {\enquote {\bibinfo {title} {Dynamics of
				structural transitions in liquids},}\ }\href@noop {} {\bibfield  {journal}
		{\bibinfo  {journal} {Phys. Rev. A}\ }\textbf {\bibinfo {volume} {28}},\
		\bibinfo {pages} {2408--2416} (\bibinfo {year} {1983})}\BibitemShut {NoStop}%
	\bibitem [{\citenamefont {Schr\o{}der}(2022)}]{sch22}%
	\BibitemOpen
	\bibfield  {author} {\bibinfo {author} {\bibfnamefont {T.~B.}\ \bibnamefont
			{Schr\o{}der}},\ }\bibfield  {title} {\enquote {\bibinfo {title} {Predicting
				scaling properties from a single fluid configuration},}\ }\href {\doibase
		10.1103/PhysRevLett.129.245501} {\bibfield  {journal} {\bibinfo  {journal}
			{Phys. Rev. Lett.}\ }\textbf {\bibinfo {volume} {129}},\ \bibinfo {pages}
		{245501} (\bibinfo {year} {2022})}\BibitemShut {NoStop}%
	\bibitem [{\citenamefont {Koperwas}\ \emph {et~al.}(2020)\citenamefont
		{Koperwas}, \citenamefont {Grzybowski},\ and\ \citenamefont
		{Paluch}}]{kop20}%
	\BibitemOpen
	\bibfield  {author} {\bibinfo {author} {\bibfnamefont {K.}~\bibnamefont
			{Koperwas}}, \bibinfo {author} {\bibfnamefont {A.}~\bibnamefont
			{Grzybowski}}, \ and\ \bibinfo {author} {\bibfnamefont {M.}~\bibnamefont
			{Paluch}},\ }\bibfield  {title} {\enquote {\bibinfo {title}
			{Virial--potential-energy correlation and its relation to density scaling for
				quasireal model systems},}\ }\href {\doibase 10.1103/PhysRevE.102.062140}
	{\bibfield  {journal} {\bibinfo  {journal} {Phys. Rev. E}\ }\textbf {\bibinfo
			{volume} {102}},\ \bibinfo {pages} {062140} (\bibinfo {year}
		{2020})}\BibitemShut {NoStop}%
	\bibitem [{\citenamefont {Sheydaafar}\ \emph {et~al.}(2023)\citenamefont
		{Sheydaafar}, \citenamefont {Dyre},\ and\ \citenamefont
		{Schr{\o}der}}]{she23}%
	\BibitemOpen
	\bibfield  {author} {\bibinfo {author} {\bibfnamefont {Z.}~\bibnamefont
			{Sheydaafar}}, \bibinfo {author} {\bibfnamefont {J.~C.}\ \bibnamefont
			{Dyre}}, \ and\ \bibinfo {author} {\bibfnamefont {T.~B.}\ \bibnamefont
			{Schr{\o}der}},\ }\bibfield  {title} {\enquote {\bibinfo {title} {Scaling
				properties of liquid dynamics predicted from a single configuration: {Small}
				rigid molecules},}\ }\href {\doibase 10.1021/acs.jpcb.3c01574} {\bibfield
		{journal} {\bibinfo  {journal} {J. Phys. Chem. B}\ }\textbf {\bibinfo
			{volume} {127}},\ \bibinfo {pages} {3478--3487} (\bibinfo {year}
		{2023})}\BibitemShut {NoStop}%
	\bibitem [{\citenamefont {Vrabec}\ \emph {et~al.}(2001)\citenamefont {Vrabec},
		\citenamefont {Stoll},\ and\ \citenamefont {Hasse}}]{vra01}%
	\BibitemOpen
	\bibfield  {author} {\bibinfo {author} {\bibfnamefont {J.}~\bibnamefont
			{Vrabec}}, \bibinfo {author} {\bibfnamefont {J.}~\bibnamefont {Stoll}}, \
		and\ \bibinfo {author} {\bibfnamefont {H.}~\bibnamefont {Hasse}},\ }\bibfield
	{title} {\enquote {\bibinfo {title} {A set of molecular models for symmetric
				quadrupolar flu},}\ }\href {\doibase 10.1021/jp012542o} {\bibfield  {journal}
		{\bibinfo  {journal} {J. Phys. Chem. B}\ }\textbf {\bibinfo {volume} {105}},\
		\bibinfo {pages} {12126--12133} (\bibinfo {year} {2001})}\BibitemShut
	{NoStop}%
	\bibitem [{\citenamefont {Galbraith}\ and\ \citenamefont {Hall}(2007)}]{gal07}%
	\BibitemOpen
	\bibfield  {author} {\bibinfo {author} {\bibfnamefont {A.~L.}\ \bibnamefont
			{Galbraith}}\ and\ \bibinfo {author} {\bibfnamefont {C.~K.}\ \bibnamefont
			{Hall}},\ }\bibfield  {title} {\enquote {\bibinfo {title} {Solid–liquid
				phase equilibria for mixtures containing diatomic {Lennard–Jones}
				molecules},}\ }\href {\doibase https://doi.org/10.1016/j.fluid.2007.07.064}
	{\bibfield  {journal} {\bibinfo  {journal} {Fluid Phase Eq.}\ }\textbf
		{\bibinfo {volume} {262}},\ \bibinfo {pages} {1--13} (\bibinfo {year}
		{2007})}\BibitemShut {NoStop}%
	\bibitem [{\citenamefont {Schr{\o}der}\ \emph
		{et~al.}(2009{\natexlab{a}})\citenamefont {Schr{\o}der}, \citenamefont
		{Bailey}, \citenamefont {Pedersen}, \citenamefont {Gnan},\ and\ \citenamefont
		{Dyre}}]{III}%
	\BibitemOpen
	\bibfield  {author} {\bibinfo {author} {\bibfnamefont {T.~B.}\ \bibnamefont
			{Schr{\o}der}}, \bibinfo {author} {\bibfnamefont {N.~P.}\ \bibnamefont
			{Bailey}}, \bibinfo {author} {\bibfnamefont {U.~R.}\ \bibnamefont
			{Pedersen}}, \bibinfo {author} {\bibfnamefont {N.}~\bibnamefont {Gnan}}, \
		and\ \bibinfo {author} {\bibfnamefont {J.~C.}\ \bibnamefont {Dyre}},\
	}\bibfield  {title} {\enquote {\bibinfo {title} {Pressure-energy correlations
				in liquids. {III. Statistical} mechanics and thermodynamics of liquids with
				hidden scale invariance},}\ }\href {\doibase 10.1063/1.3265955} {\bibfield
		{journal} {\bibinfo  {journal} {J. Chem. Phys.}\ }\textbf {\bibinfo {volume}
			{131}},\ \bibinfo {pages} {234503} (\bibinfo {year}
		{2009}{\natexlab{a}})}\BibitemShut {NoStop}%
	\bibitem [{\citenamefont {Chopra}\ \emph
		{et~al.}(2010{\natexlab{a}})\citenamefont {Chopra}, \citenamefont
		{Truskett},\ and\ \citenamefont {Errington}}]{cho10a}%
	\BibitemOpen
	\bibfield  {author} {\bibinfo {author} {\bibfnamefont {R.}~\bibnamefont
			{Chopra}}, \bibinfo {author} {\bibfnamefont {T.~M.}\ \bibnamefont
			{Truskett}}, \ and\ \bibinfo {author} {\bibfnamefont {J.~R.}\ \bibnamefont
			{Errington}},\ }\bibfield  {title} {\enquote {\bibinfo {title}
			{Excess-entropy scaling of dynamics for a confined fluid of dumbbell-shaped
				particles},}\ }\href@noop {} {\bibfield  {journal} {\bibinfo  {journal}
			{Phys. Rev. E}\ }\textbf {\bibinfo {volume} {82}},\ \bibinfo {pages} {041201}
		(\bibinfo {year} {2010}{\natexlab{a}})}\BibitemShut {NoStop}%
	\bibitem [{\citenamefont {Chopra}\ \emph
		{et~al.}(2010{\natexlab{b}})\citenamefont {Chopra}, \citenamefont
		{Truskett},\ and\ \citenamefont {Errington}}]{cho10b}%
	\BibitemOpen
	\bibfield  {author} {\bibinfo {author} {\bibfnamefont {R.}~\bibnamefont
			{Chopra}}, \bibinfo {author} {\bibfnamefont {T.~M.}\ \bibnamefont
			{Truskett}}, \ and\ \bibinfo {author} {\bibfnamefont {J.~R.}\ \bibnamefont
			{Errington}},\ }\bibfield  {title} {\enquote {\bibinfo {title} {Excess
				entropy scaling of dynamic quantities for fluids of dumbbell-shaped
				particles},}\ }\href@noop {} {\bibfield  {journal} {\bibinfo  {journal} {J.
				Chem. Phys.}\ }\textbf {\bibinfo {volume} {133}},\ \bibinfo {pages} {104506}
		(\bibinfo {year} {2010}{\natexlab{b}})}\BibitemShut {NoStop}%
	\bibitem [{\citenamefont {Fragiadakis}\ and\ \citenamefont
		{Roland}(2017)}]{fra17}%
	\BibitemOpen
	\bibfield  {author} {\bibinfo {author} {\bibfnamefont {D.}~\bibnamefont
			{Fragiadakis}}\ and\ \bibinfo {author} {\bibfnamefont {C.~M}\ \bibnamefont
			{Roland}},\ }\bibfield  {title} {\enquote {\bibinfo {title} {A test for the
				existence of isomorphs in glass-forming materials},}\ }\href {\doibase
		10.1063/1.4986774} {\bibfield  {journal} {\bibinfo  {journal} {J. Chem.
				Phys.}\ }\textbf {\bibinfo {volume} {147}},\ \bibinfo {pages} {084508}
		(\bibinfo {year} {2017})}\BibitemShut {NoStop}%
	\bibitem [{\citenamefont {Santiago}(2018)}]{san18}%
	\BibitemOpen
	\bibfield  {author} {\bibinfo {author} {\bibfnamefont {I.}~\bibnamefont
			{Santiago}},\ }\bibfield  {title} {\enquote {\bibinfo {title} {Nanoscale
				active matter matters: {Challenges} and opportunities for self-propelled
				nanomotors},}\ }\href {\doibase https://doi.org/10.1016/j.nantod.2018.01.001}
	{\bibfield  {journal} {\bibinfo  {journal} {Nano Today}\ }\textbf {\bibinfo
			{volume} {19}},\ \bibinfo {pages} {11--15} (\bibinfo {year}
		{2018})}\BibitemShut {NoStop}%
	\bibitem [{\citenamefont {Dombrowski}\ and\ \citenamefont
		{Klotsa}(2020)}]{dom20}%
	\BibitemOpen
	\bibfield  {author} {\bibinfo {author} {\bibfnamefont {T.}~\bibnamefont
			{Dombrowski}}\ and\ \bibinfo {author} {\bibfnamefont {D.}~\bibnamefont
			{Klotsa}},\ }\bibfield  {title} {\enquote {\bibinfo {title} {Kinematics of a
				simple reciprocal model swimmer at intermediate {Reynolds} numbers},}\ }\href
	{\doibase 10.1103/PhysRevFluids.5.063103} {\bibfield  {journal} {\bibinfo
			{journal} {Phys. Rev. Fluids}\ }\textbf {\bibinfo {volume} {5}},\ \bibinfo
		{pages} {063103} (\bibinfo {year} {2020})}\BibitemShut {NoStop}%
	\bibitem [{\citenamefont {Schr{\o}der}\ \emph
		{et~al.}(2009{\natexlab{b}})\citenamefont {Schr{\o}der}, \citenamefont
		{Pedersen}, \citenamefont {Bailey}, \citenamefont {Toxvaerd},\ and\
		\citenamefont {Dyre}}]{sch09}%
	\BibitemOpen
	\bibfield  {author} {\bibinfo {author} {\bibfnamefont {T.~B.}\ \bibnamefont
			{Schr{\o}der}}, \bibinfo {author} {\bibfnamefont {U.~R.}\ \bibnamefont
			{Pedersen}}, \bibinfo {author} {\bibfnamefont {N.~P.}\ \bibnamefont
			{Bailey}}, \bibinfo {author} {\bibfnamefont {S.}~\bibnamefont {Toxvaerd}}, \
		and\ \bibinfo {author} {\bibfnamefont {J.~C.}\ \bibnamefont {Dyre}},\
	}\bibfield  {title} {\enquote {\bibinfo {title} {{Hidden Scale Invariance in
					Molecular van der Waals Liquids: A Simulation Study}},}\ }\href@noop {}
	{\bibfield  {journal} {\bibinfo  {journal} {Phys. Rev. E}\ }\textbf {\bibinfo
			{volume} {80}},\ \bibinfo {pages} {041502} (\bibinfo {year}
		{2009}{\natexlab{b}})}\BibitemShut {NoStop}%
	\bibitem [{\citenamefont {Milinkovic}\ \emph {et~al.}(2013)\citenamefont
		{Milinkovic}, \citenamefont {Dennison},\ and\ \citenamefont
		{Dijkstra}}]{mil13}%
	\BibitemOpen
	\bibfield  {author} {\bibinfo {author} {\bibfnamefont {K.}~\bibnamefont
			{Milinkovic}}, \bibinfo {author} {\bibfnamefont {M.}~\bibnamefont
			{Dennison}}, \ and\ \bibinfo {author} {\bibfnamefont {M.}~\bibnamefont
			{Dijkstra}},\ }\bibfield  {title} {\enquote {\bibinfo {title} {Phase diagram
				of hard asymmetric dumbbell particles},}\ }\href {\doibase
		10.1103/PhysRevE.87.032128} {\bibfield  {journal} {\bibinfo  {journal} {Phys.
				Rev. E}\ }\textbf {\bibinfo {volume} {87}},\ \bibinfo {pages} {032128}
		(\bibinfo {year} {2013})}\BibitemShut {NoStop}%
	\bibitem [{\citenamefont {Bennemann}\ \emph {et~al.}(1998)\citenamefont
		{Bennemann}, \citenamefont {Paul}, \citenamefont {Binder},\ and\
		\citenamefont {D{\"u}nweg}}]{ben98}%
	\BibitemOpen
	\bibfield  {author} {\bibinfo {author} {\bibfnamefont {C.}~\bibnamefont
			{Bennemann}}, \bibinfo {author} {\bibfnamefont {W.}~\bibnamefont {Paul}},
		\bibinfo {author} {\bibfnamefont {K.}~\bibnamefont {Binder}}, \ and\ \bibinfo
		{author} {\bibfnamefont {B.}~\bibnamefont {D{\"u}nweg}},\ }\bibfield  {title}
	{\enquote {\bibinfo {title} {Molecular-dynamics simulations of the thermal
				glass transition in polymer melts: \ensuremath{\alpha}-relaxation
				behavior},}\ }\href {\doibase 10.1103/PhysRevE.57.843} {\bibfield  {journal}
		{\bibinfo  {journal} {Phys. Rev. E}\ }\textbf {\bibinfo {volume} {57}},\
		\bibinfo {pages} {843--851} (\bibinfo {year} {1998})}\BibitemShut {NoStop}%
	\bibitem [{\citenamefont {Aichele}\ \emph {et~al.}(2003)\citenamefont
		{Aichele}, \citenamefont {Gebremichael}, \citenamefont {Starr}, \citenamefont
		{Baschnagel},\ and\ \citenamefont {Glotzer}}]{aic03}%
	\BibitemOpen
	\bibfield  {author} {\bibinfo {author} {\bibfnamefont {M.}~\bibnamefont
			{Aichele}}, \bibinfo {author} {\bibfnamefont {Y.}~\bibnamefont
			{Gebremichael}}, \bibinfo {author} {\bibfnamefont {F.~W.}\ \bibnamefont
			{Starr}}, \bibinfo {author} {\bibfnamefont {J.}~\bibnamefont {Baschnagel}}, \
		and\ \bibinfo {author} {\bibfnamefont {S.~C.}\ \bibnamefont {Glotzer}},\
	}\bibfield  {title} {\enquote {\bibinfo {title} {Polymer-specific effects of
				bulk relaxation and stringlike correlated motion in the dynamics of a
				supercooled polymer melt},}\ }\href {\doibase 10.1063/1.1597473} {\bibfield
		{journal} {\bibinfo  {journal} {J. Chem. Phys.}\ }\textbf {\bibinfo {volume}
			{119}},\ \bibinfo {pages} {5290--5304} (\bibinfo {year} {2003})}\BibitemShut
	{NoStop}%
	\bibitem [{\citenamefont {Puosi}\ and\ \citenamefont {Leporini}(2011)}]{puo11}%
	\BibitemOpen
	\bibfield  {author} {\bibinfo {author} {\bibfnamefont {F.}~\bibnamefont
			{Puosi}}\ and\ \bibinfo {author} {\bibfnamefont {D.}~\bibnamefont
			{Leporini}},\ }\bibfield  {title} {\enquote {\bibinfo {title} {Scaling
				between relaxation, transport, and caged dynamics in polymers: {From} cage
				restructuring to diffusion},}\ }\href {\doibase 10.1021/jp203659r} {\bibfield
		{journal} {\bibinfo  {journal} {J. Phys. Chem. B}\ }\textbf {\bibinfo
			{volume} {115}},\ \bibinfo {pages} {14046--14051} (\bibinfo {year}
		{2011})}\BibitemShut {NoStop}%
	\bibitem [{\citenamefont {Shavit}\ \emph {et~al.}(2013)\citenamefont {Shavit},
		\citenamefont {Douglas},\ and\ \citenamefont {Riggleman}}]{sha13}%
	\BibitemOpen
	\bibfield  {author} {\bibinfo {author} {\bibfnamefont {A.}~\bibnamefont
			{Shavit}}, \bibinfo {author} {\bibfnamefont {J.~F.}\ \bibnamefont {Douglas}},
		\ and\ \bibinfo {author} {\bibfnamefont {R.~A.}\ \bibnamefont {Riggleman}},\
	}\bibfield  {title} {\enquote {\bibinfo {title} {Evolution of collective
				motion in a model glass-forming liquid during physical aging},}\ }\href
	{\doibase 10.1063/1.4775781} {\bibfield  {journal} {\bibinfo  {journal} {J.
				Chem. Phys.}\ }\textbf {\bibinfo {volume} {138}},\ \bibinfo {pages} {12A528}
		(\bibinfo {year} {2013})}\BibitemShut {NoStop}%
	\bibitem [{\citenamefont {Bailey}\ \emph {et~al.}(2017)\citenamefont {Bailey},
		\citenamefont {Ingebrigtsen}, \citenamefont {Hansen}, \citenamefont
		{Veldhorst}, \citenamefont {B{\o}hling}, \citenamefont {Lemarchand},
		\citenamefont {Olsen}, \citenamefont {Bacher}, \citenamefont {Costigliola},
		\citenamefont {Pedersen}, \citenamefont {Larsen}, \citenamefont {Dyre},\ and\
		\citenamefont {Schr{\o}der}}]{RUMD}%
	\BibitemOpen
	\bibfield  {author} {\bibinfo {author} {\bibfnamefont {N.~P.}\ \bibnamefont
			{Bailey}}, \bibinfo {author} {\bibfnamefont {T.~S.}\ \bibnamefont
			{Ingebrigtsen}}, \bibinfo {author} {\bibfnamefont {J.~S.}\ \bibnamefont
			{Hansen}}, \bibinfo {author} {\bibfnamefont {A.~A.}\ \bibnamefont
			{Veldhorst}}, \bibinfo {author} {\bibfnamefont {L.}~\bibnamefont
			{B{\o}hling}}, \bibinfo {author} {\bibfnamefont {C.~A.}\ \bibnamefont
			{Lemarchand}}, \bibinfo {author} {\bibfnamefont {A.~E.}\ \bibnamefont
			{Olsen}}, \bibinfo {author} {\bibfnamefont {A.~K.}\ \bibnamefont {Bacher}},
		\bibinfo {author} {\bibfnamefont {L.}~\bibnamefont {Costigliola}}, \bibinfo
		{author} {\bibfnamefont {U.~R.}\ \bibnamefont {Pedersen}},\ 
		{\bibnamefont {et al.}},\ }\bibfield  {title} {\bibinfo
		{title} {{RUMD}: A general purpose molecular dynamics package optimized to
			utilize {GPU} hardware down to a few thousand particles},\ }\href@noop {}
	{\bibfield  {journal} {\bibinfo  {journal} {Scipost Phys.}\ }\textbf
		{\bibinfo {volume} {3}},\ \bibinfo {pages} {038} (\bibinfo {year}
		{2017})}\BibitemShut {NoStop}
	\bibitem [{\citenamefont {Dyre}(2018{\natexlab{a}})}]{dyr18a}%
	\BibitemOpen
	\bibfield  {author} {\bibinfo {author} {\bibfnamefont {J.~C.}\ \bibnamefont
			{Dyre}},\ }\bibfield  {title} {\enquote {\bibinfo {title} {Perspective:
				Excess-entropy scaling},}\ }\href {\doibase 10.1063/1.5055064} 
			{\bibfield	{journal} {\bibinfo  {journal} {J. Chem. Phys.}\ }
				\textbf {\bibinfo {volume}	{149}},\ 
				\bibinfo {pages} {210901} (\bibinfo {year}
		{2018}{\natexlab{a}})}\BibitemShut {NoStop}%
	\bibitem [{\citenamefont {Rosenfeld}(1977)}]{ros77}%
	\BibitemOpen
	\bibfield  {author} {\bibinfo {author} {\bibfnamefont {Y.}~\bibnamefont
			{Rosenfeld}},\ }\bibfield  {title} {\enquote {\bibinfo {title} {Relation
				between the transport coefficients and the internal entropy of simple
				systems},}\ }\href@noop {} {\bibfield  {journal} {\bibinfo  {journal} {Phys.
				Rev. A}\ }\textbf {\bibinfo {volume} {15}},\ \bibinfo {pages} {2545--2549}
		(\bibinfo {year} {1977})}\BibitemShut {NoStop}%
	\bibitem [{\citenamefont {Dyre}(2018{\natexlab{b}})}]{dyr18}%
	\BibitemOpen
	\bibfield  {author} {\bibinfo {author} {\bibfnamefont {J.~C.}\ \bibnamefont
			{Dyre}},\ }\bibfield  {title} {\enquote {\bibinfo {title} {Isomorph theory of
				physical aging},}\ }\href {\doibase 10.1063/1.5022999} {\bibfield  {journal}
		{\bibinfo  {journal} {J. Chem. Phys.}\ }\textbf {\bibinfo {volume} {148}},\
		\bibinfo {pages} {154502} (\bibinfo {year} {2018}{\natexlab{b}})}\BibitemShut
	{NoStop}%
	\bibitem [{\citenamefont {Alba-Simionesco}\ \emph {et~al.}(2004)\citenamefont
		{Alba-Simionesco}, \citenamefont {Cailliaux}, \citenamefont {Alegria},\ and\
		\citenamefont {Tarjus}}]{alb04}%
	\BibitemOpen
	\bibfield  {author} {\bibinfo {author} {\bibfnamefont {C.}~\bibnamefont
			{Alba-Simionesco}}, \bibinfo {author} {\bibfnamefont {A.}~\bibnamefont
			{Cailliaux}}, \bibinfo {author} {\bibfnamefont {A.}~\bibnamefont {Alegria}},
		\ and\ \bibinfo {author} {\bibfnamefont {G.}~\bibnamefont {Tarjus}},\
	}\bibfield  {title} {\enquote {\bibinfo {title} {Scaling out the density
				dependence of the alpha relaxation in glass-forming polymers},}\ }\href
	{\doibase 10.1209/epl/i2004-10214-6} {\bibfield  {journal} {\bibinfo
			{journal} {Europhys. Lett.}\ }\textbf {\bibinfo {volume} {68}},\ \bibinfo
		{pages} {58--64} (\bibinfo {year} {2004})}\BibitemShut {NoStop}%
	\bibitem [{\citenamefont {Roland}\ \emph {et~al.}(2005)\citenamefont {Roland},
		\citenamefont {Hensel-Bielowka}, \citenamefont {Paluch},\ and\ \citenamefont
		{Casalini}}]{rol05}%
	\BibitemOpen
	\bibfield  {author} {\bibinfo {author} {\bibfnamefont {C.~M.}\ \bibnamefont
			{Roland}}, \bibinfo {author} {\bibfnamefont {S.}~\bibnamefont
			{Hensel-Bielowka}}, \bibinfo {author} {\bibfnamefont {M.}~\bibnamefont
			{Paluch}}, \ and\ \bibinfo {author} {\bibfnamefont {R.}~\bibnamefont
			{Casalini}},\ }\bibfield  {title} {\enquote {\bibinfo {title} {Supercooled
				dynamics of glass-forming liquids and polymers under hydrostatic pressure},}\
	}\href {\doibase 10.1088/0034-4885/68/6/R03} {\bibfield  {journal} {\bibinfo
			{journal} {Rep. Prog. Phys.}\ }\textbf {\bibinfo {volume} {68}},\ \bibinfo
		{pages} {1405--1478} (\bibinfo {year} {2005})}\BibitemShut {NoStop}%
	\bibitem [{\citenamefont {Roland}\ \emph {et~al.}(2003)\citenamefont {Roland},
		\citenamefont {Casalini},\ and\ \citenamefont {Paluch}}]{rol03}%
	\BibitemOpen
	\bibfield  {author} {\bibinfo {author} {\bibfnamefont {C.~M.}\ \bibnamefont
			{Roland}}, \bibinfo {author} {\bibfnamefont {R.}~\bibnamefont {Casalini}}, \
		and\ \bibinfo {author} {\bibfnamefont {M.}~\bibnamefont {Paluch}},\
	}\bibfield  {title} {\enquote {\bibinfo {title} {Isochronal
				temperature--pressure superpositioning of the alpha--relaxation in type-{A}
				glass formers},}\ }\href@noop {} {\bibfield  {journal} {\bibinfo  {journal}
			{Chem. Phys. Lett.}\ }\textbf {\bibinfo {volume} {367}},\ \bibinfo {pages}
		{259--264} (\bibinfo {year} {2003})}\BibitemShut {NoStop}%
	\BibitemShut {NoStop}%
	\bibitem [{\citenamefont {Knudsen}\ \emph {et~al.}(2021)\citenamefont
		{Knudsen}, \citenamefont {Niss},\ and\ \citenamefont {Bailey}}]{knu21}%
	\BibitemOpen	
	\bibfield  {author} {\bibinfo {author} {\bibfnamefont {P.~A.}\ \bibnamefont
			{Knudsen}}, \bibinfo {author} {\bibfnamefont {K.}~\bibnamefont {Niss}}, \
		and\ \bibinfo {author} {\bibfnamefont {N.~P.}\ \bibnamefont {Bailey}},\
	}\bibfield  {title} {\enquote {\bibinfo {title} {Quantifying dynamical and
				structural invariance in a simple molten salt model},}\ }\href {\doibase
		10.1063/5.0055794} {\bibfield  {journal} {\bibinfo  {journal} {J. Chem.
				Phys.}\ }\textbf {\bibinfo {volume} {155}},\ \bibinfo {pages} {054506}
		(\bibinfo {year} {2021})}\BibitemShut {NoStop}%
	\bibitem [{\citenamefont {Knudsen}\ \emph {et~al.}(2024)\citenamefont
		{Knudsen}, \citenamefont {Heyes}, \citenamefont {Niss}, \citenamefont
		{Dini},\ and\ \citenamefont {Bailey}}]{knu24}%
	\BibitemOpen
	\bibfield  {author} {\bibinfo {author} {\bibfnamefont {P.~A.}\ \bibnamefont
			{Knudsen}}, \bibinfo {author} {\bibfnamefont {D.~M.}\ \bibnamefont {Heyes}},
		\bibinfo {author} {\bibfnamefont {K.}~\bibnamefont {Niss}}, \bibinfo {author}
		{\bibfnamefont {D.}~\bibnamefont {Dini}}, \ and\ \bibinfo {author}
		{\bibfnamefont {N.~P.}\ \bibnamefont {Bailey}},\ }\bibfield  {title}
	{\enquote {\bibinfo {title} {Invariant dynamics in a united-atom model of an
				ionic liquid},}\ }\href {\doibase 10.1063/5.0177373} {\bibfield  {journal}
		{\bibinfo  {journal} {J. Chem. Phys.}\ }\textbf {\bibinfo {volume} {160}},\
		\bibinfo {pages} {034503} (\bibinfo {year} {2024})}\BibitemShut {NoStop}%
\end{thebibliography}
\end{document}